\documentclass[12pt]{elsarticle} 

\usepackage{amsmath,amsthm}
\usepackage{blkarray}
\usepackage{amsfonts,amssymb}
\usepackage{url}
\usepackage{mathtools}  
\usepackage[colorlinks=true,urlcolor=blue,linkcolor=red,citecolor=magenta]{hyperref}
\usepackage{enumerate, paralist}
\usepackage{tikz-cd}
\usepackage{xcolor}
\usepackage{hyperref}
\usepackage{soul}
\usepackage{subfig}
\usepackage{cleveref} 
\usepackage{pgfplots}
\pgfplotsset{compat=1.18}
\usepackage{multirow}
\usepackage{longtable}
\usepackage{array}

\begin{document}

\begin{frontmatter}

\title{What is missing from this picture? Persistent homology and mixup barcodes as a means of investigating negative embedding space}

\author[1]{Himanshu Yadav}
\ead{yadav.himanshu@ufl.edu}

\author[2]{Thomas Bryan Smith}
\ead{tbsmit10@olemiss.edu}

\author[1]{Peter Bubenik}
\ead{peter.bubenik@ufl.edu}

\author[3]{Christopher McCarty}
\ead{ufchris@ufl.edu}

\affiliation[1]{organization={Department of Mathematics, University of Florida},%Department and Organization
            %addressline={}, 
            city={Gainesville},
            postcode={32611}, 
            state={Florida},
            country={USA}}

\affiliation[2]{organization={Department of Criminal Justice and Legal Studies, University of Mississippi},%Department and Organization
            %addressline={}, 
            city={Oxford},
            postcode={38677}, 
            state={Mississippi},
            country={USA}}

\affiliation[3]{organization={Bureau of Economic and Business Research, University of Florida},%Department and Organization
            %addressline={}, 
            city={Gainesville},
            postcode={32611}, 
            state={Florida},
            country={USA}}

% -------------------------------------------------------------------%

\begin{abstract}
Recent work in the information sciences, especially informetrics and scientometrics, has made substantial contributions to the development of new metrics that eschew the intrinsic biases of citation metrics. 
This work has tended to employ either network scientific (topological) approaches to quantifying the disruptiveness of peer-reviewed research, or topic modeling approaches to quantifying conceptual novelty. 
We propose a combination of these approaches, investigating the prospect of topological data analysis (TDA), specifically persistent homology and mixup barcodes, as a means of understanding the negative space among document embeddings generated by topic models. 
Using \textit{top2vec}, we embed documents and topics in \textit{n}-dimensional space, we use persistent homology to identify `holes' in the embedding distribution, and then use mixup barcodes to determine which holes are being filled by a set of unobserved publications.
In this case, the unobserved publications represent research that was published before or after the data used to train \textit{top2vec}. 
We investigate the extent that negative embedding space represents missing context (older research) versus innovation space (newer research), and the extend that the documents that occupy this space represents integrations of the research topics on the periphery.
Potential applications for this metric are discussed.
\end{abstract}

\begin{keyword}
Persistent homology \sep
Mixup barcodes \sep
Topological data analysis (TDA) \sep
Science of Science (SciSci)\sep

\end{keyword}

\end{frontmatter}

\section{Introduction}

Science policy and innovation studies have long pursued methods for quantifying and studying innovation and scientific impact. \cite{MARTIN20121219, SutherlandEtAl2011, SinatraEtAl2016}.
Produced primarily by scholars working within sociology of knowledge and information sciences (scientometrics,  informetrics, among others), much of the research in this area replicates and extends analyses of individual productivity, established innovation metrics (e.g., citation impact), and their predictors \cite{Andersonetal2004}.  
However, the recent advancement of a `Science of Science' (SciSci) as an emerging subfield has coincided with a flurry of advancements in the measurement of innovation and disruption \cite{Wu_Wang_Evans_2019, Hofstra_et_al_2020}. 
SciSci is a transdisciplinary approach that leverages data science and advanced computational methods to study the mechanisms of `doing science' \cite{Fortunatoetal2018}. 
Recent research has utilized topological (network) methods to distinguish disruptive and developing research. 
This is achieved by constructing directed acyclic graphs of academic research citation networks and measuring the tendency of scholars to cite disruptive research excusively (original articles, nobel prize work, etc.), but cite developing articles (e.g., review articles) in tandem with preceding work \cite{Wu_Wang_Evans_2019}. 
Subsequent research further supports this notion, quantifying innovation as both the introduction of new concepts and the novel combination of existing concepts (e.g., cross-disciplinary integration) \cite{Hofstra_et_al_2020, Leahey2023}.
This research, among other work, highlight a growing concern with the development of new scientific innovation metrics that eschew the intrinsic biases of citation metrics by accounting for conceptual novelty.
For example, citation metrics have long been recognized as problematic for the arts, humanities, and social sciences, where books and book chapters remain a normal outlet for knowledge dissemination \cite{Thelwall_et_al_2015}.
Since neither books or book chapters appear in the Web of Science, they do not feature in the calculation of many impact factor and similar metrics \cite{Diaz-Faes_et_al_2016, Bornmann_et_al_2016}.
We investigate the prospect of topological data analysis (TDA)\cite{munch2017user}, including persistent homology \cite{edelsbrunner2010computational} and mixup barcodes \cite{Wagner:2024aa}, as a means of understanding the negative space among document embeddings and the prospect that documents occupying this space represent innovative, interdisciplinary scholarship.

\subsection{Word and Document Embedding}

Responding to the rapidly increasing growth of scientific knowledge and a perceived challenges of `information overload' \cite{Landhuis_2016}, SciSci scholars have begun "Science Mapping", the application of techniques from network science and, more recently, computational linguistics to scientific literature with the goal of summarizing and understanding the `landscape' of a given field \cite{Chen2017140}. Earlier research in this area has focused heavily on the construction, visualization, and interpretation of citation and coauthorship networks \cite{ARIA2017959}. However, rapid development in natural language processing (NLP) and machine learning have afforded scholars a wealth of text summarization algorithms, many of which rely on word and document embeddings.
Conceptually, word and document embedding (and topic modeling, more broadly) can be understood as an assortative process wherein each document set (corpus) and the words therein are placed in \textit{n}-dimensional space such that substantively similar documents and words are proximal, while dissimilar documents are distal \cite{Salton_68, angelov2020, grootendorst2022}. 
Thus, to some extent, word and document embeddings can be understood as a conceptual landscape. 

\subsection{Structural linguistics and conceptual landscapes}

There are some notable limitations in representing words, documents, and topics as embeddings or `conceptual landscapes.'
The meaning of text is not completely represented by a relational framework arising from traditional structural linguistics \cite{saussure1983course}.
Some words are `embodied' in that they represent sensory information, others might vary substantially based on the context of the author (i.e., each word is not universally coherent), and many words are not static over time \cite{Arseniev-Koehler2024}. 
Absent the capacity to `embody' information, given that language models do not have the capacity to understand the words themselves, it is important that we investigate the concepts and meaning that embeddings fail to \textit{directly} capture.
Practically speaking, information and meaning are lost when a language model is trained on data collected from a specific point or period of time, creating a truncated training set.
Missing data that precede the first training data point might represent potentially important missing context, while missing data beyond the last training data point might more frequently represent innovative, novel recombinations of established ideas \cite{Weitzman1998, Leahey&Moody2014}.
To use an analogy, let us say that we have a disassembled jigsaw puzzle with missing pieces and duplicate pieces, and each piece of the jigsaw puzzle is a document. 
Topic embedding models assemble the puzzle, presenting us with the complete `picture' of the conceptual landscape, which then allows us to (1) identify dense clusters of duplicate puzzle pieces, and (2) identify incomplete areas of the puzzle. 
Topic models are only intended to identify dense clusters; they effectively disregard negative space.
We propose the application of topological data analysis (TDA) - specifically persistent homology and mixup barcodes - as a means of examining the negative space amid embedded words, documents, and topics.
In other words, we propose the application of TDA to identify and examine the missing puzzle pieces.
Just as TDA has been used to map \textit{geographical} space in geographical information science \cite{Corcoran04032023}, there is potential for similar applications in \textit{conceptual} space, moving beyond one constraint imposed by a rigid structural linguistic framework.

\subsection{Applications of Topological Data Analysis}

TDA has been applied in various fields and subfields, including but not limited to geographic information systems \cite{Feng_2021}, resource allocation \cite{Hickok_2024}, neuroscience \cite{Thomas_2021}, environmental science \cite{Ver_Hoef_2023}, and material science \cite{Hiraoka_2016}. 
Non-theoretical applications of topology have proven valuable to many domains of science, and one method that has been particularly valuable is TDA \cite{DONUT}. 
Among these applications, TDA has been used to find topological features for high-dimensional data and then find lower-dimensional projections preserving these structures \cite{Wang_2011, Perea2023, McInnes2020}. 
Due to the efficacy of TDA tools to deal with high-dimensional data, some research has employed TDA in the study of topological structures in textual data, per the aforementioned structural linguistic underpinnings of work in this domain \cite{Uchendu_2024}. 
Recently, Dragnaov et al. \cite{Draganov_2024} used TDA to demonstrate that topological features for word embeddings for different languages vary significantly. 
Gholizadeh et al. \cite{gholizadeh2020} also found that topological structure in word embeddings on long textual documents provides a more accurate linguistic representation than conventional text mining features.
This work, among others, showcases the interesting topological features that underpin word embeddings.
It follows that TDA techniques designed to identify and examine the birth, death, and mixup of cavities could also help to examine and better understand the negative embedding space among embedded words, documents, and topics.

\section{Background}

\subsection{Top2vec}
\textit{Top2vec} is an extension of \textit{word2vec} and \textit{doc2vec} that applies dimension reduction and hierarchical clustering to word and document embeddings with the goal of detecting coherent themes and topics in a given body of text \cite{angelov2020}.
The underlying models, \textit{word2vec} and \textit{doc2vec}, are trained using the distributed bag-of-words (DBOW) architecture, with the goal of generating joint embeddings for words and documents \cite{le2014}. 
As such, \textit{top2vec} takes the input word matrix $W_{n,d}$ and the word context matrix $W'_{n,d}$, then uses each word vector $\overrightarrow{w}\in W_{n,d}$ for word $w$ to predict the context vector $\overrightarrow{w}\in W'_{n,d}$ of a given context word $w_{c}$. 
With the $softmax(\overrightarrow{w}\in W'_{n,d})$ activation function, backpropagation, and stochastic gradient descent, \textit{ top2vec} then estimates the probability distribution of a given context vector conditional on a given word vector, $P( \overrightarrow{w_{c}} | \overrightarrow{w} )$ \cite{angelov2020}. 
This process results in a set of $k$ embeddings (i.e. the context vectors $W'$) which are closer together for semantically similar words, and distant for semantically dissimilar words.

Per the DBOW \textit{doc2vec} framework, \textit{top2vec} learns document embeddings by following a very similar process, but taking the document matrix $D_{c,d}$ as input, rather than a word matrix, producing an equivalent set of $k$ embeddings that are proximal for documents containing similar words, and distal for documents containing dissimilar words \cite{le2014,angelov2020}.
Taking jointly embedded words and documents as input, uniform manifold approximation and projection (UMAP) is used for dimension reduction while preserving the local and global structure of the embeddings \cite{angelov2020}. 
UMAP uses components of TDA to generate fuzzy topological representations of embedded words and documents, and finds a set of low-dimensional embeddings that produce an approximately equivalent fuzzy topological representation \cite{McInnes2020}. 
Following dimension reduction, hierarchical density-based clustering (HDBSCAN) is applied to the reduced embeddings, identifying sufficiently dense clusters of reduced document vectors while eliminating noise \cite{McInnes2017,angelov2020, Campelloetal2015}.
The result is a set of embedded words and documents that have been clustered into a collection of topics that represent coherent themes in the corpus.

\subsection{Persistent homology}

Topological data analysis follows the premise that the shape of data (and the conceptual, spatial information that encodes this shape) contains substantive information.
Assuming that the shape of the conceptual landscape inferred by \textit{top2vec} is meaningful, it follows that the negative space between clusters of words, documents, and topics is meaningful.
We thus aim to identify persistent holes in the conceptual landscape using \textit{persistent homology} \cite{edelsbrunner2010computational}.

Spatial information is encoded as homology groups \cite{munkres2018elements} in topology. 
The homology group for dimension $k$ informally describes the number of $k$-dimensional holes ($k$-dimensional cycles which are not boundary).
The rank of the $k^{th}$ homology group is also the $k^{th}$ Betti number.
An element of the $k^{th}$ homology group is often called the degree-$k$ topological feature.
Homology groups capture topological information for a single choice of scale,
whereas persistent homology captures the multi-scale topological structure of the data.

\begin{figure}
    \centering
    \includegraphics[width=0.9\linewidth]{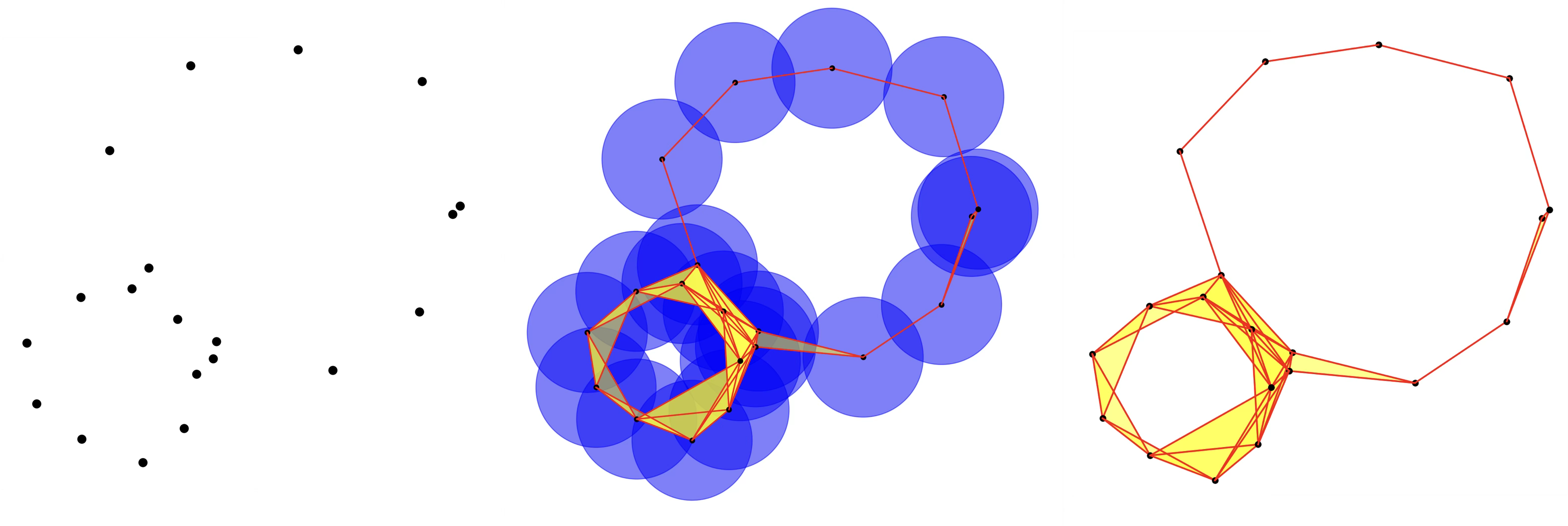}
    \caption{
    Left: the collection of black dots is called a point cloud.
    Middle: consider balls of a fixed radius centered at each point.
    If two balls intersect, join them by an edge.
    If three balls have a common intersection,  fill the triangle between the three edges arising from the pairwise intersections.
    Similarly, add higher-dimensional simplices for  higher-order intersections.
%  Balls acts as cover for the space represented by union of the balls.
    Right: the simplicial complex (consisting of vertices, edges, triangles, and higher-order simplices) obtained from this construction.
    }
    \label{fig:nerve}
\end{figure}

The main idea of persistent homology is to track topological features as they appear and disappear as we build up our geometric object through a filtration. 
Given a point-cloud data set in an ambient space with a distance metric, we can recover the topology with the help of growing balls around each data point.
For a fixed radius, the collections of balls is a cover of the union of balls. 
Then, per the nerve theorem \cite{Borsuk1948}, the nerve of this cover is homotopy-equivalent to the union of the cover.
This guarantees that the homology of the nerve and the union of the ball are both the same.
As calculating homology of the nerve is much simpler than finding homology of a general space, this extra construction reduces computational cost.
The nerve of a cover gives a combinatorial representation of the union of balls, called a \textit{simplicial complex}.
This is the geometric object which we use for our data, see \cref{fig:nerve}.

\begin{figure}
    \subfloat[\centering Persistent homology and barcode]{
        \includegraphics[width=0.9\linewidth]{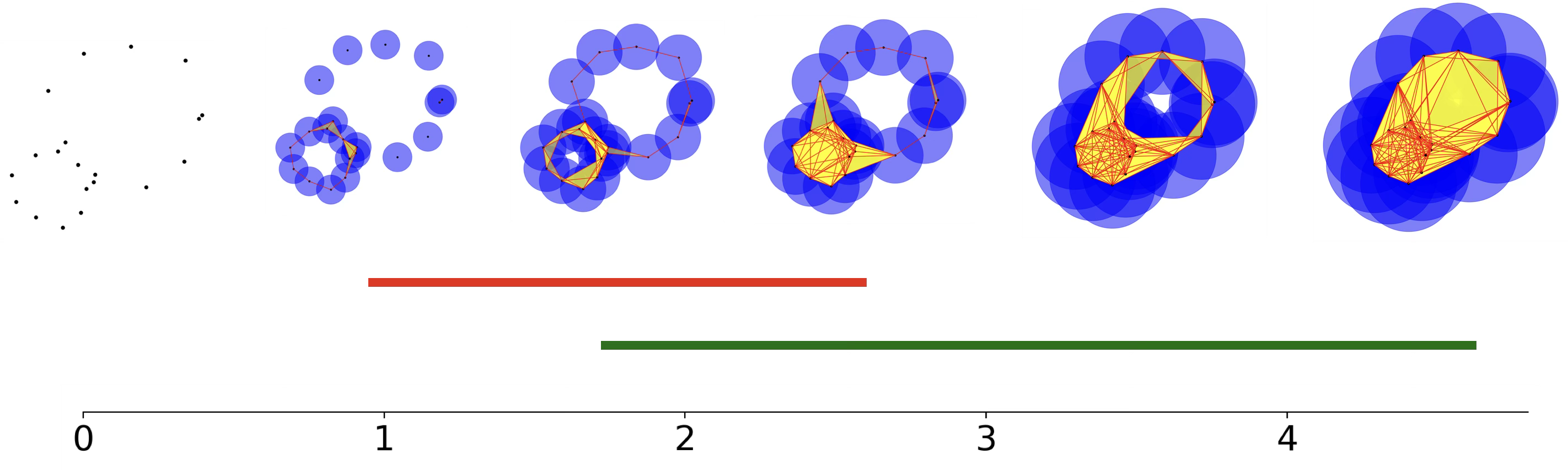}
        \label{subfig:filtration}
    }
 
    \subfloat[\centering Persistence Diagram]{
        \begin{tikzpicture}
            \begin{axis}[
                axis lines = middle,
                xlabel = {birth},
                ylabel = {death},
                xmin=0, xmax=5,
                ymin=0, ymax=5,
                xtick={0,1,2,3,4,5},
                ytick={0,1,2,3,4,5},
                width=0.45\linewidth,
                xlabel style={at={(axis description cs:0.5,-0.1)},anchor=north},
                ylabel style={rotate=-90, at={(axis description cs:-0.2,0.5)},anchor=south}
            ]   
                % Add the diagonal x=y line
                \addplot[domain=0:4.5, samples=2, color=black] {x};
                
                \addplot[
                  only marks,
                  mark=*,
                  mark options={scale=1, fill=red}
                ] coordinates {
                    (0.9,2.6)
                };

                \addplot[
                  only marks,
                  mark=*,
                  mark options={scale=1, fill=teal}
                ] coordinates {
                    (1.7,4.6)
                };
            \end{axis}
        %\draw (-1.25,-1.25) rectangle (6.25,5.25);
        \end{tikzpicture}
    }%
    \subfloat[\centering Representative cycle]{{\includegraphics[width=0.4\linewidth]{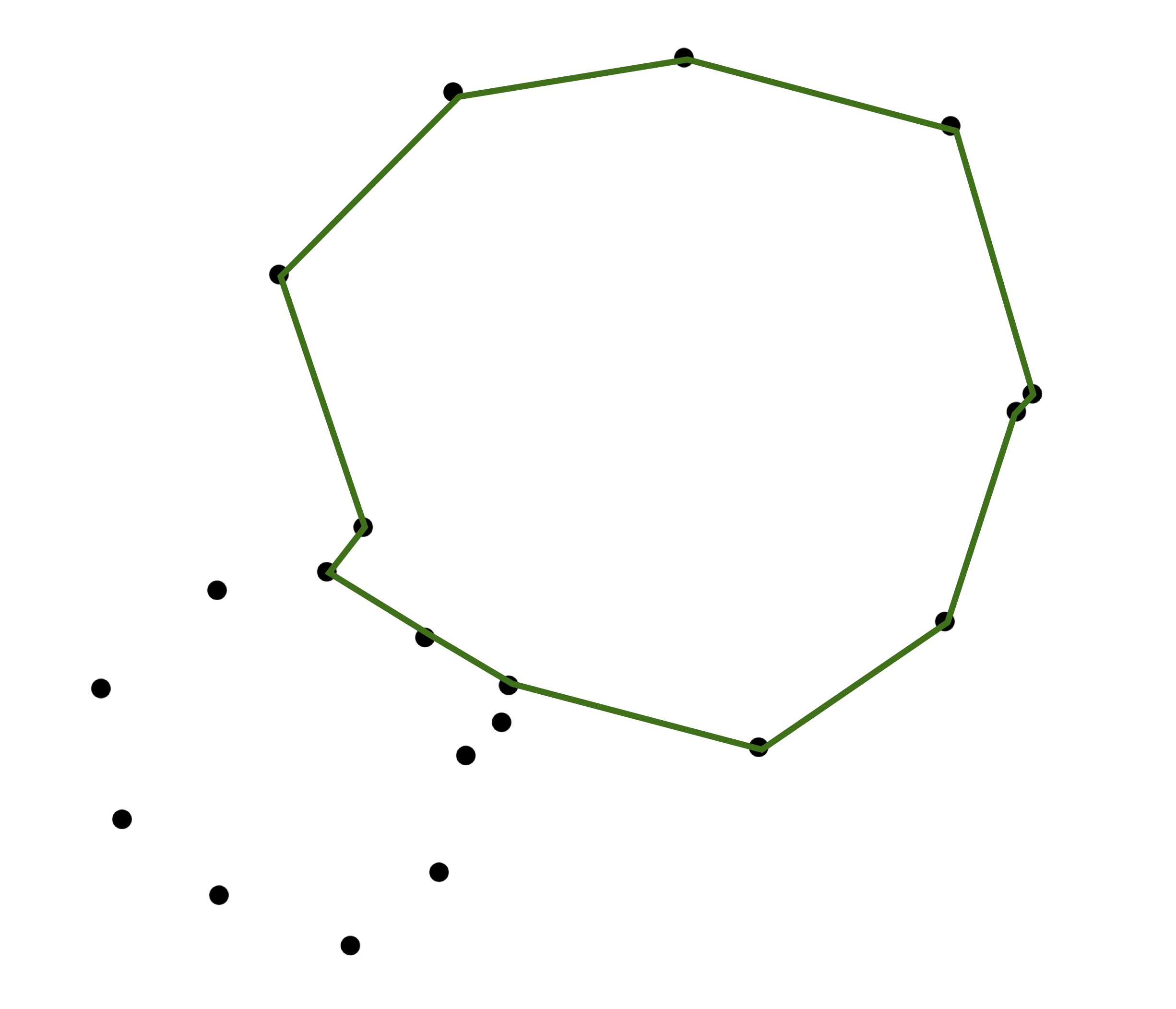}}}
    
    \caption{In (a), we recovered the homological information of a point cloud data which resembles an $8$-figure with persistent homology.
    The two colored bars correspond to two holes surrounded by a $1$-dimensional boundary.
    In (b), the persistence barcode is represented as a multi-set of points called a persistence diagram. 
    For the green point in the persistence barcode, a representative cycle is visualized as the green cycle in (c).}
    \label{barcode}
\end{figure}
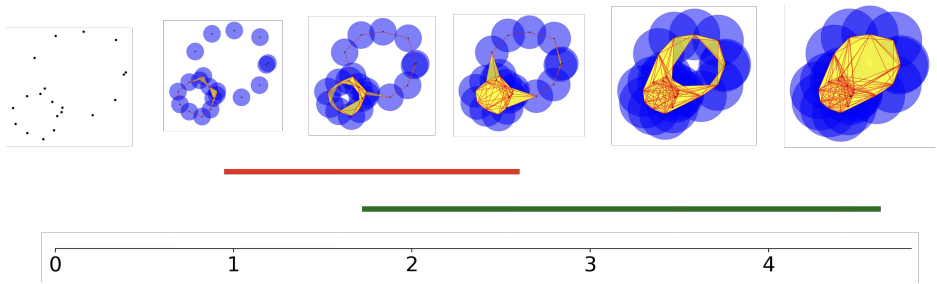
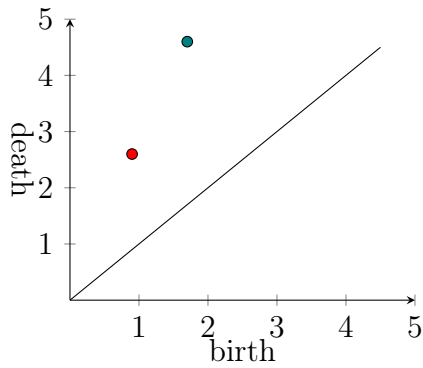
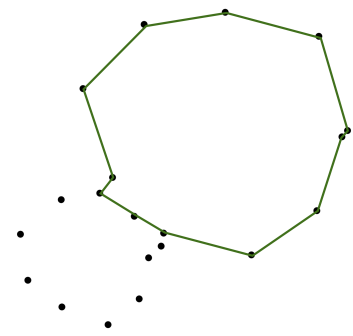

We construct a simplicial complex for a sequence of increasing radii to get a filtered simplicial complex \cref{subfig:filtration}. 
Then we take the homology of these filtered simplicial complexes with coefficients in $\mathbb{Z}_{2}$, this gives us a \textit{persistence module}.
A persistence module consists of homology groups for every step, and induced maps between two consecutive steps.
We can track topological features across these homological groups using the images and kernels of these induced maps to precisely encode which features are born, which die, and which persist.

Per structure theorem \cite{botnan2020decomposition},
one can decompose persistence modules and represent them as persistence barcodes. 
Every bar in the persistence barcode has a starting and ending value called \textit{birth} and \textit{death} respectively.
These values correspond to when a topological feature is born and dies.
We define \textit{persistence} as the difference of death and birth values,
which captures how long a feature persists in the filtration. 
Each topological feature which persists across multiple scales can be represented by a \textit{representative cycle}, a concrete geometric object that exemplifies the feature.
The birth value for a topological feature corresponds to the step when a representative cycle first emerges,
this occurs when the last simplex which forms the structure of the cycle emerges.
The death value corresponds to the step when this representative cycle becomes a boundary of a higher dimensional simplex, due to the addition of a new simplex.
The new simplex whose addition allows representative cycle to become a boundary, is called the \textit{death simplex}.

For instance, take a 1-dimensional hole in the data. 
This is represented by a loop that bounds the hole; this is its representative cycle.
The largest weight of the edges (1-dimensional simplex) in the representative cycle is birth, when the representative cycle completely appears in the filtration.
This representative cycle is later filled by a combination of triangles (2-dimensional simplices);
the largest weighted triangle is the death simplex, which appears in the end to fill the hole.

We plot birth-death in the $x$-$y$ plane as a multi-set of points called a \textit{persistence diagram}.
In \cref{barcode} we use persistent homology to recover the $1$st homology group for the point cloud which resembles the figure eight. 
In \cref{barcode} (c), we visualize a $1$-dimensional hole by its representative cycle.
We will only compute persistent homology for dimension one in this paper.

\subsection{Mixup Barcodes}\label{sec_mixup}
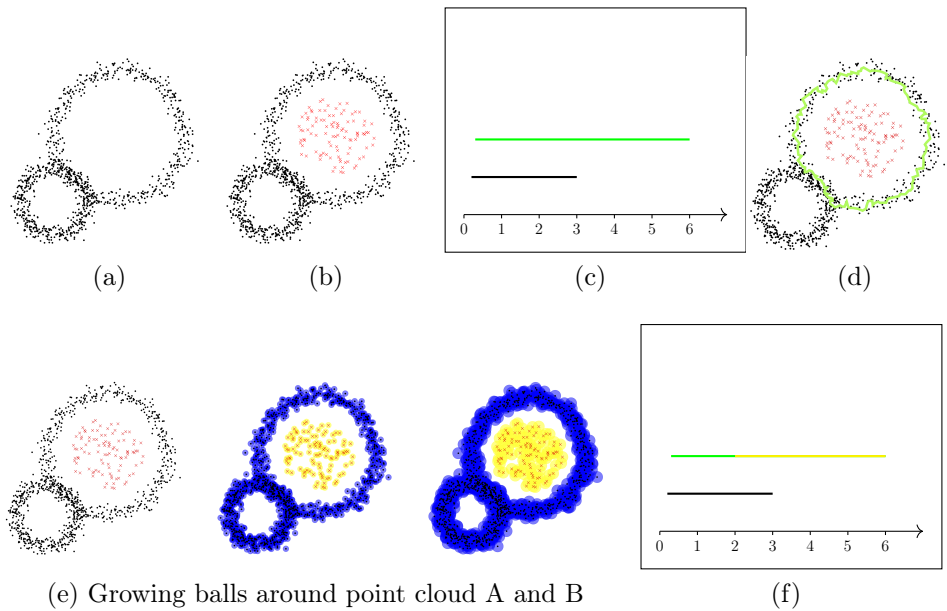
\begin{figure}%
    \subfloat[]{{\includegraphics[width=0.2\linewidth]{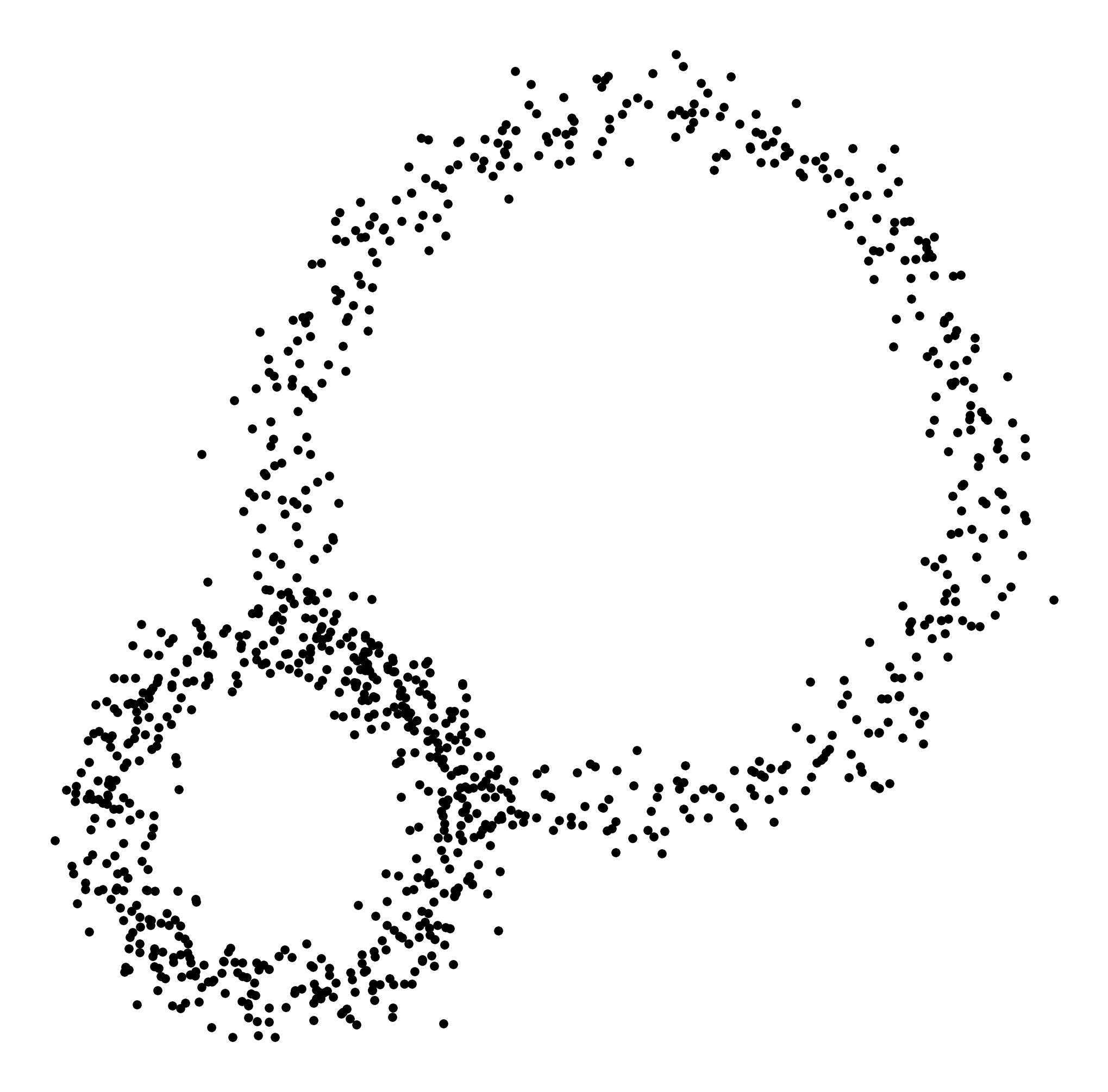} }}%
    \subfloat[]{{\includegraphics[width=0.2\linewidth]{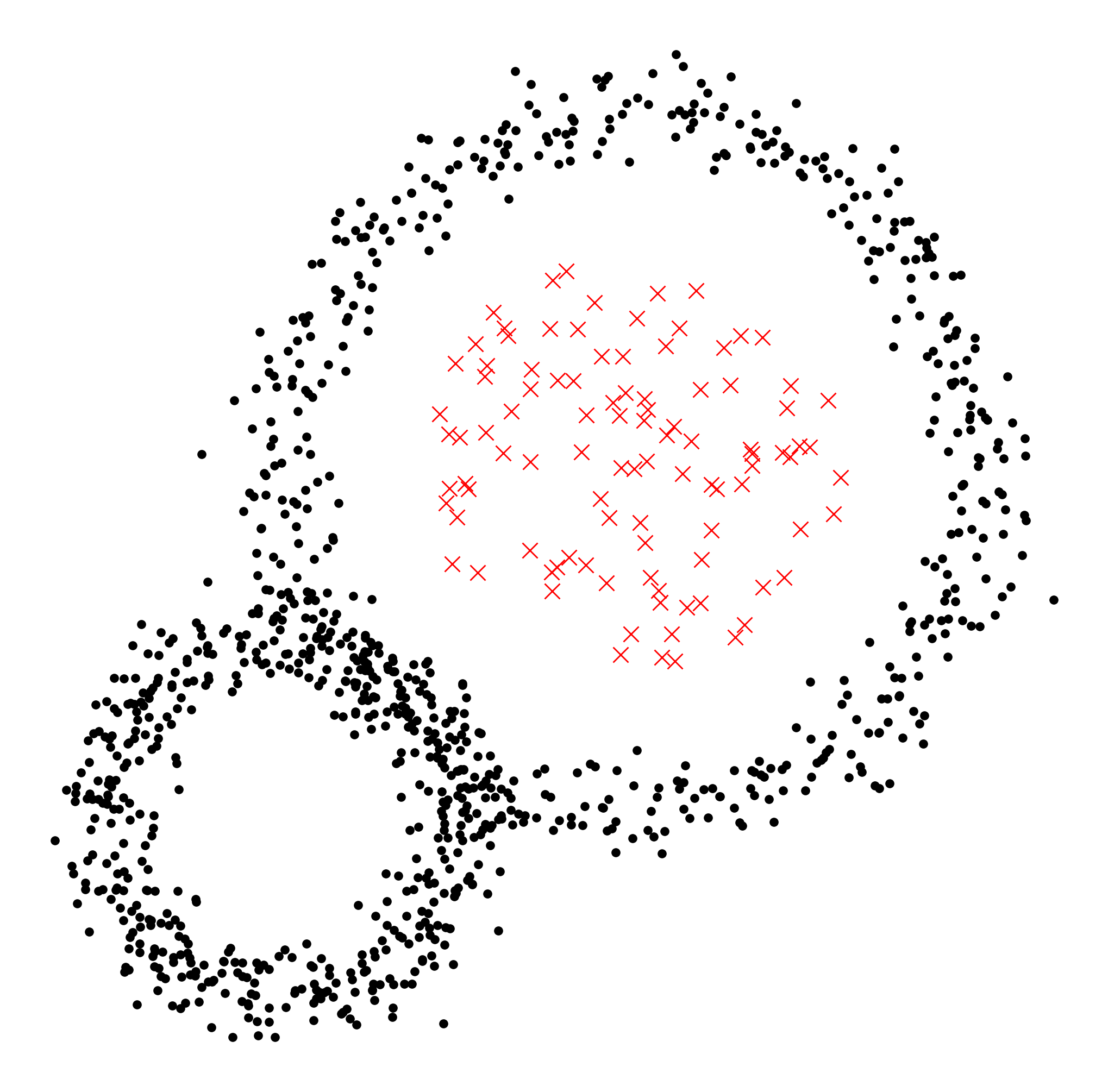} }}%
    \subfloat[]{
    \begin{tikzpicture}[scale=0.5, transform shape]
    % Draw x and y axes
    \draw[->] (0,0) -- (7,0) node[right] {};
    
    % Draw the first line from x=1 to x=4
    \draw[thick, black] (0.2,1) -- (3,1);
    
    % Draw the second line from x=2 to x=6
    \draw[thick, green] (0.3,2) -- (6,2);
    
    % Add labels for the x-coordinates
    \foreach \x in {0,1,2,3,4,5,6}
        \draw (\x,0) -- (\x,-0.1) node[below] {\x};
    
    % Add square box around the plot
    \draw (-0.5,-1) rectangle (7.5,5.5);
    \end{tikzpicture}}
    \subfloat[]{{\includegraphics[width=0.2\linewidth]{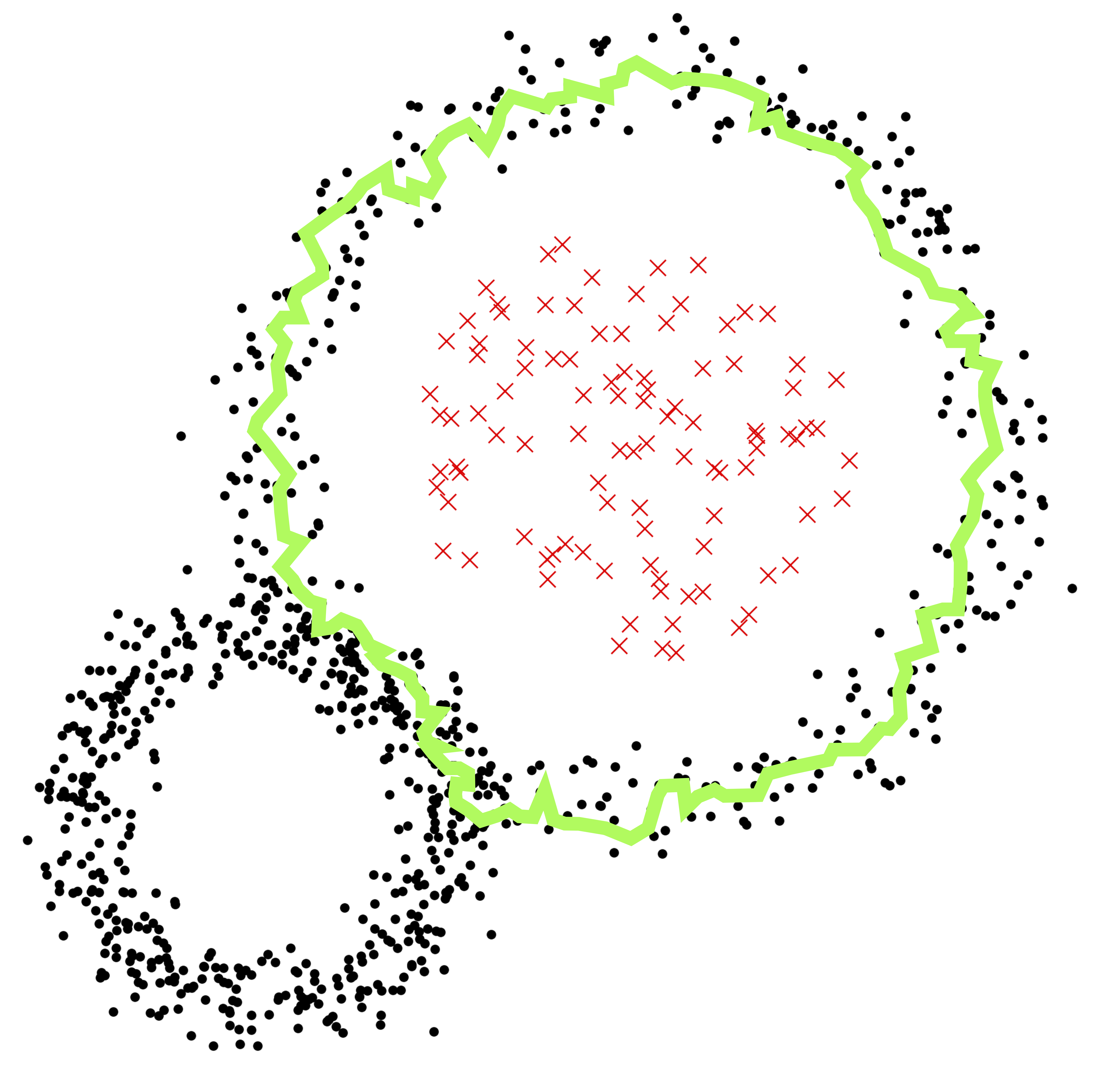} }}
    %\qquad
    \qquad
    
    \subfloat[\centering Growing balls around point cloud A and B]{{\includegraphics[width=0.6\linewidth]{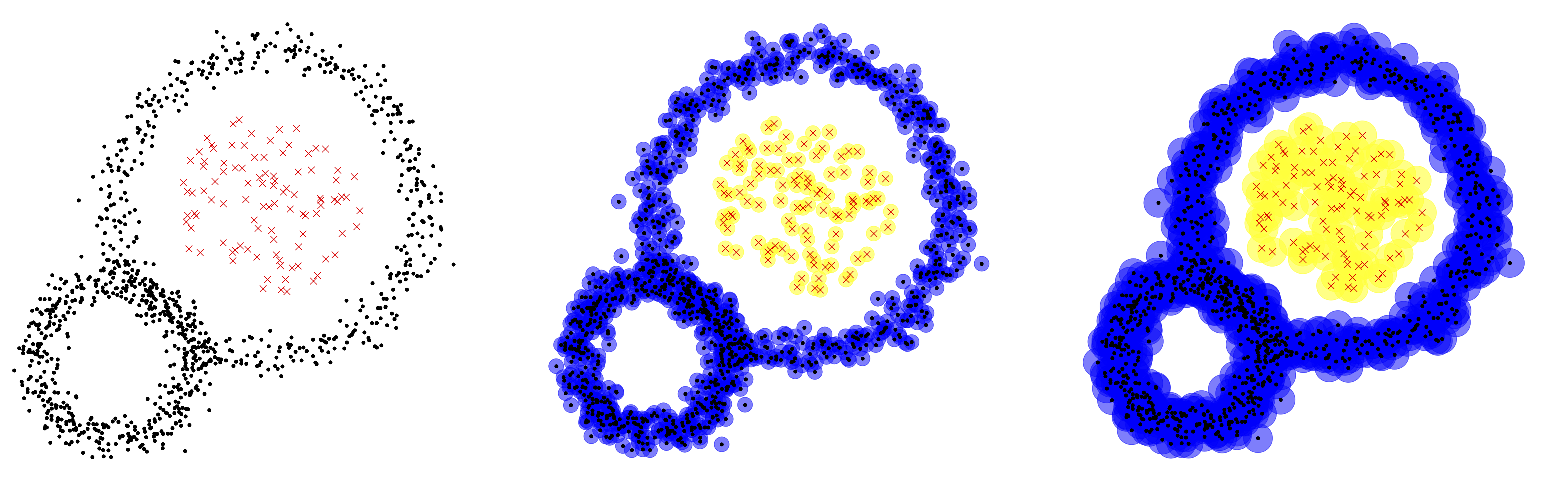} }}
    \subfloat[]{
    \begin{tikzpicture}[scale=0.5, transform shape]
    % Draw x and y axes
    \draw[->] (0,0) -- (7,0) node[right] {};
    
    % Draw the first line from x=1 to x=4
    \draw[thick, black] (0.2,1) -- (3,1);
    
    % Draw the second line from x=2 to x=6
    \draw[thick, green] (0.3,2) -- (6,2);
    % Draw the second line from x=2 to x=6
    \draw[thick, yellow] (2,2) -- (6,2);
    
    % Add labels for the x-coordinates
    \foreach \x in {0,1,2,3,4,5,6}
        \draw (\x,0) -- (\x,-0.1) node[below] {\x};
    
    % Add square box around the plot
    \draw (-0.5,-1) rectangle (7.5,5.5);
    \end{tikzpicture}}

    \caption{Point cloud $P$ contains the black dots, and point cloud $Q$ contains the red crosses. 
    (a) Point cloud $P$.
    (b) Point clouds $P$ and $Q$.
    Using standard persistent homology we can find holes in dimension $1$ for point cloud $P$. 
    In (c), persistence barcode with the two longest bars for the $1$st homology group of point cloud $P$ is visualized. 
    In (d), representative cycle of the longest bar (green color) is visualized in point cloud $P$ and $Q$. 
    Now, we consider how the inclusion of point cloud $Q$ effect the 
    hole visualized by the representative cycle in green color.
    (e) Growing balls around point cloud $P$ and $Q$ fills the hole visualized by the green representative cycle earlier compared to when we just have point cloud $P$.
    (f) The yellow bar represents how the inclusion of point cloud $Q$ into point cloud $P$ has reduced the length of the original green bar.}%
    
    \label{mixup}%
\end{figure}

Standard persistent homology captures the topology of a point cloud which has one group or classification of data points, say from a distribution $P$.
If we add more points into this point cloud from a different distribution $Q$, the $k$-dimensional holes made by points $P$ in some cases will fill up earlier because of points in $Q$.
In \cref{mixup}, an example of a topological interaction is visualized in which a $1$-dimensional hole for the point cloud $P$ fills up earlier after inclusion of points $Q$. 
This is the interaction that we use to capture using \textit{mixup barcodes}, a summary statistic developed by Wagner et al. \cite{Wagner:2024aa} to quantify topological interactions between two different points in a point cloud.
Mixup barcodes are computed using \textit{image persistence} and standard persistent homology.

First, using standard persistent homology for the point cloud $P$, birth and death simplices are identified for topological features.
Then, using image persistence,
we track when topological features which appear in the point cloud $P$ die when these new simplices from the point cloud $P\cup Q$ are included into the filtration as well. 
We will call this simplex from the point cloud $P\cup Q$ – which kills a topological feature in the point cloud $P$ – the \textit{mixup simplex}.
This mixup simplex for a topological feature will always have filtration value less than or equal to the filtration value for the corresponding death simplex obtained from the point cloud $P$.
Thus, for a homological feature in the point cloud $P$,
we can obtain birth, mixup, and death simplex with filtration values, say $b$, $d^{'}$, and $d$ respectively,
such that $b\leq d^{'}\leq d$.
If $d^{'}=d$, then adding additional points from the distribution $Q$ doesn't have any effect on the corresponding topological feature for the point cloud $P$.
However, if $d^{'}<d$, then it means that the homological feature was filled up earlier when points from the distribution $Q$ were included in the point cloud.
A statistic which can capture this interaction for a homological feature is the \textbf{mixup} given by $\frac{d-d^{'}}{d-b}$.
The \textbf{total mixup} for $k$-dimensional homology is defined as the sum of the mixup for all features in $k$-dimensional homology. 
We will only use total mixup for the $1$-dimensional homology group in the analysis of our data.
We calculate mixup using Ripserer \cite{Cufar2020}, which is a Julia library.

\section{Method}

\subsection{Data}
We used \textit{dimensions.ai} publication data, which contained every peer-reviewed article published between 2001 and 2023 by researchers indicated to be affiliated with the University of Florida (UF) circa 2023.
The data include the author's name and \textit{dimensions.ai} identifier, the university department and college, the discipline, the year of publication, and the abstract of the article.
These data were downloaded using the \textit{dimensions.ai} API via \textit{rdimensions}, an R package developed by the UF Clinical and Translational Science Institute (CTSI) Network Science team \cite{Krenz2023tilltnet}.
Publications with missing abstracts were removed from the data prior to model training.
The final dataset consisted of $114822$ publications indexed by \textit{dimensions.ai} with a minimum of one UF-affiliated author.

\subsection{\textit{Top2vec} Specification}
Given the goal of analyzing the conceptual landscape (and negative space) produced by research at the UF, we used \textit{top2vec} to construct an embedding space based on the aforementioned data.
This embedding space approximates the conceptual landscape.
Using a proximity measure (cosine similarity), \textit{top2vec} was used to further categorize publications into topics, obtaining 692 clusters of publications. 
These mutually disjoint clusters represent the broader research topics within which each publication is categorized.
This was intended to condense the corpus into a more concise representation of the conceptual landscape that can be analyzed using TDA techniques. 

The algorithms underlying \textit{top2vec} require the specification of a number of hyperparameters.
We opted to follow the conventions established by Angelov (2020) when training the model \cite{angelov2020}.
\textit{Doc2vec} requires the specification of \textit{vector size} (aka \textit{embedding dimension}) and \textit{window size} hyper-parameters - these hyperparameters were set to 300 and 15 respectively, with hierarchical softmax but no negative sampling.
UMAP requires the specification of the $n$ \textit{nearest neighbors} and \textit{embedding dimension} hyper-parameters - these were set to 50 and 5, respectively.
HDBSCAN requires the specification of a \textit{minimum cluster size} hyper-parameter, which was set to 15.

\begin{figure}%
    \subfloat[\centering]{{\includegraphics[width=0.4\linewidth]{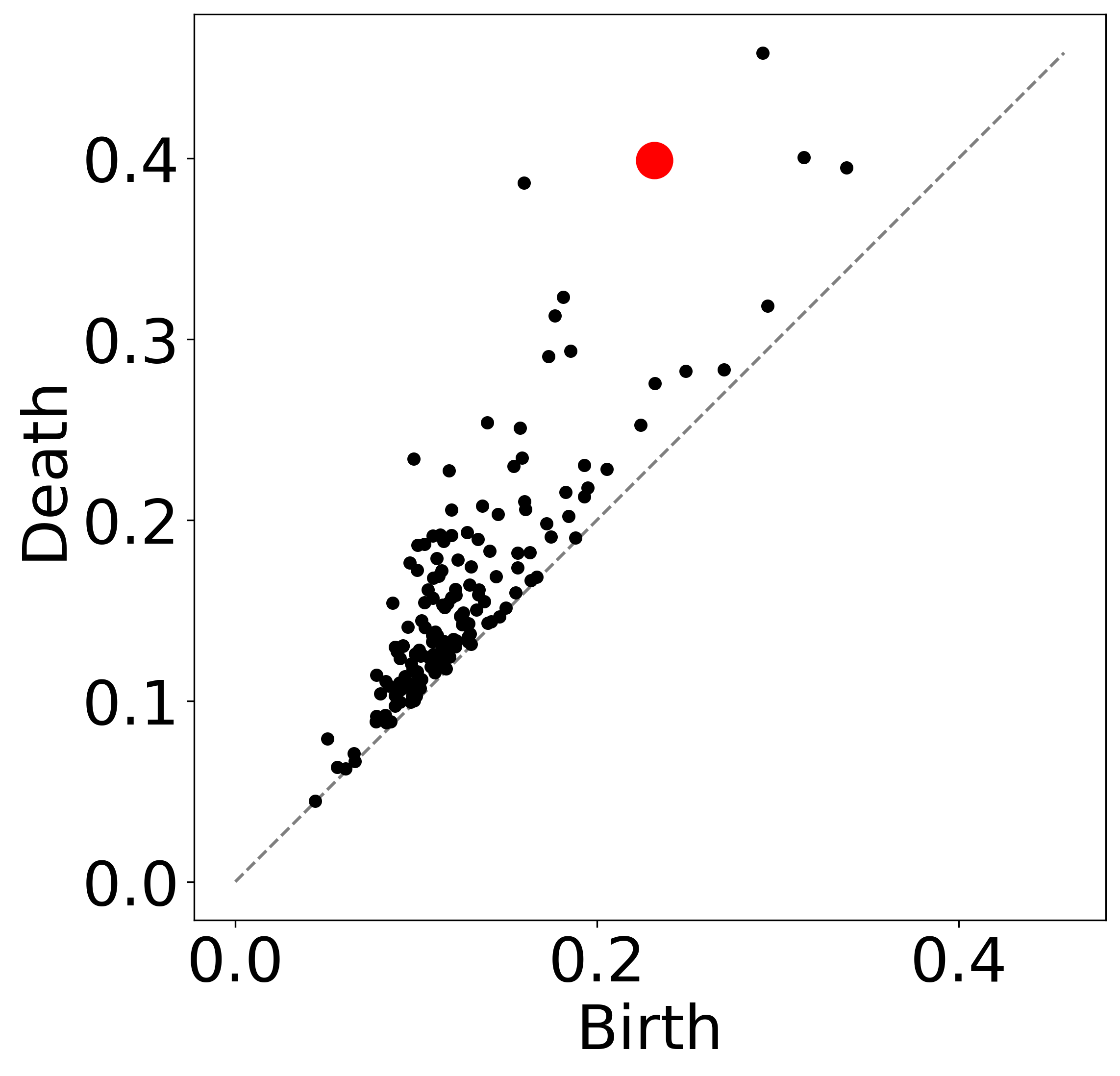} }}%
    \subfloat[\centering]{{\includegraphics[width=0.5\linewidth]{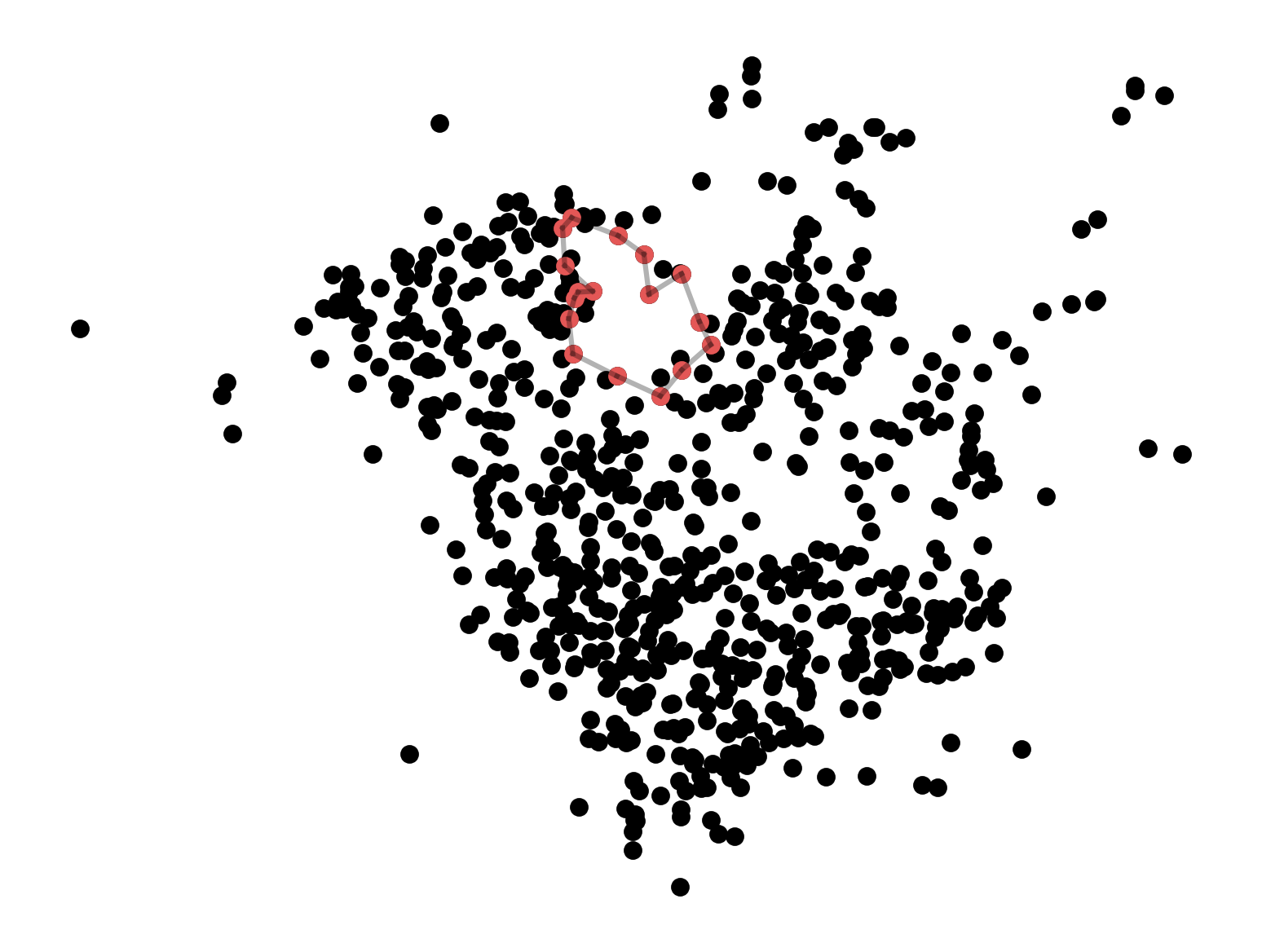} 
    }\label{subfig:topic_embedding}}%
    
    \subfloat[\centering]{{\includegraphics[width=0.4\linewidth]{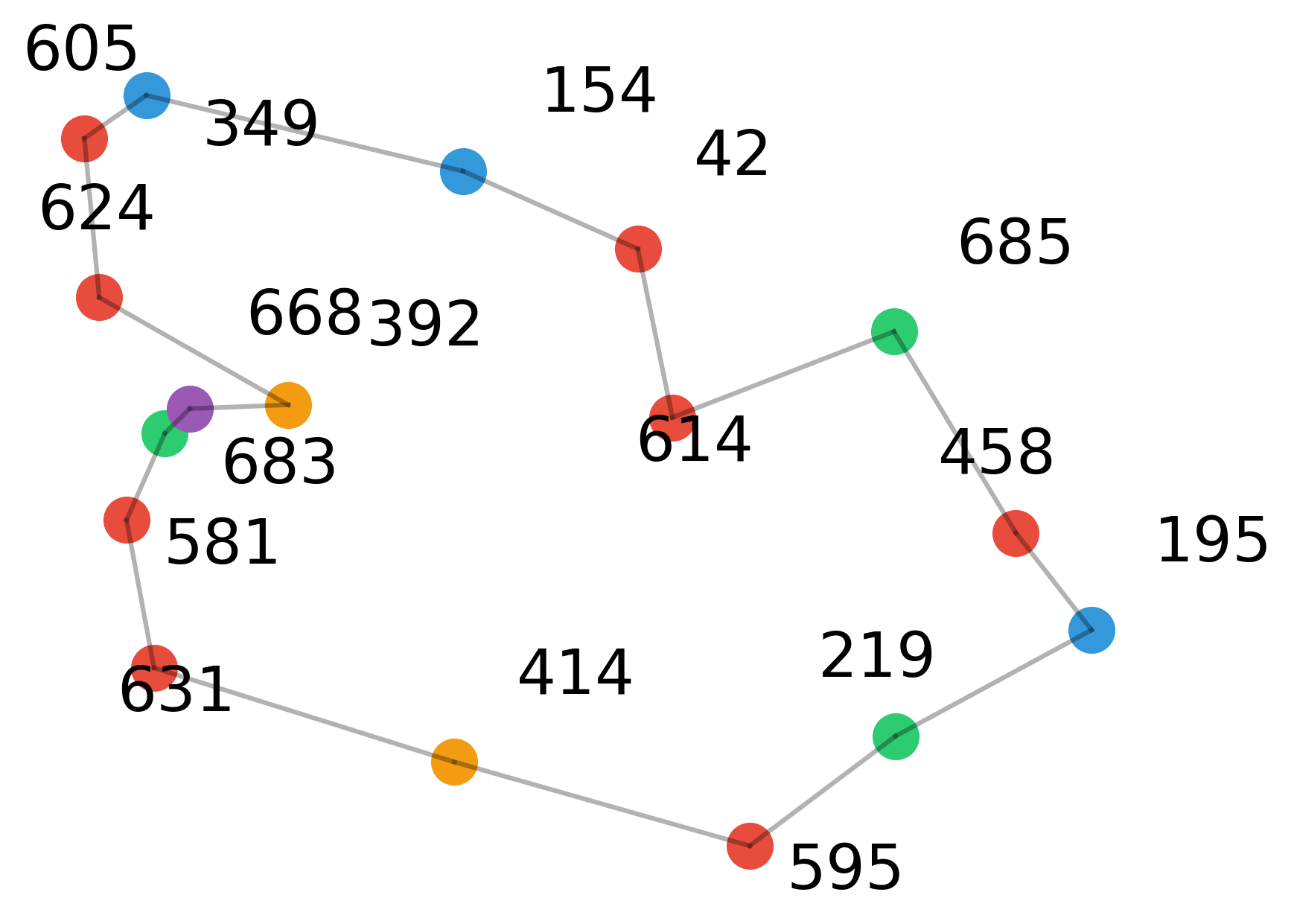} }}%
    \subfloat[\centering]{{\includegraphics[width=0.4\linewidth]{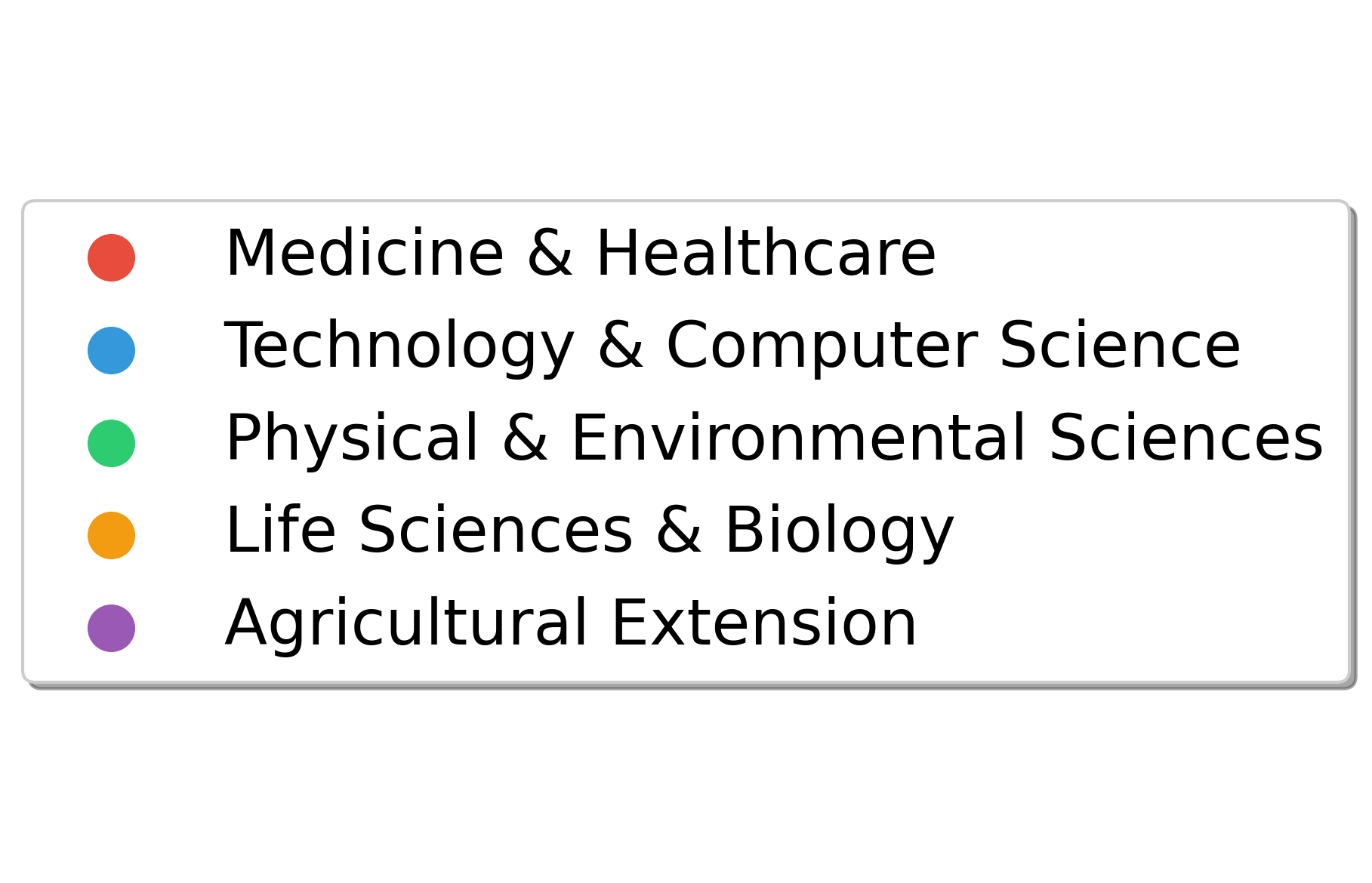} }}%
    
    \subfloat[\centering]{
      \tiny
      \begin{tabular}{|c|l|c|l|}
        \hline
        \textbf{Index} & \textbf{Topic} & \textbf{Index} & \textbf{Topic} \\
        \hline
        42 & Dentistry/Dental Care & 595 & Psychology/Counseling \\
        \hline
        154 & Cybersecurity/Computer Security & 605 & Anesthesiology/Developmental Neuroscience \\
        \hline
        195 & Mobile Health/Digital Health & 614 & Rheumatology/Musculoskeletal Medicine \\
        \hline
        219 & Environmental Biogeochemistry & 624 & HIV/AIDS Research \\
        \hline
        349 & Social Media/Digital Communication & 631 & Respiratory Physiology \\
        \hline
        392 & Olfaction/Smell Research & 668 & Agricultural Extension/Publications \\
        \hline
        414 & Insect Behavior/Pest Control & 683 & Atmospheric Physics/Lightning Research \\
        \hline
        458 & Neuroscience/Synaptic Function & 685 & Meteorology/Disaster Resilience \\
        \hline
        581 & Vascular Surgery/Cardiovascular Medicine & & \\
        \hline
        \end{tabular}
    }%
    
    \caption{
     (a) Persistence diagram in degree $1$ with a feature highlighted in red for topics embeddings obtained after projection into 2d using UMAP.
     (b) Extracted representative cycles plotted on the topics embedding for the highlighted feature in red. 
     (c) Representative cycle with topic index.
     (d) Subject classification for different topics. 
     (e) Topic information for topic indices which are part of the representative cycle.
        }%
    \label{pd_topic_embed_152}%
\end{figure}

\subsection{Analytical Procedure}
Our data lies in $5$ dimensions after projection by UMAP.
We will do our analysis in $5$ dimensions.
We present $2$-dimensional approximations alongside our 5-dimensional findings for ease of interpretation.
In \cref{subfig:topic_embedding}, the $2$-dimensional projection is visualized,
where topics were not uniformly distributed in the conceptual space. 
There were negative spaces with regions of variable densities.
Our goal was to understand this negative space under the assumption that the shape of the data contains substantive information.
In order to understand these negative spaces, which we refer to as "holes" or "cavities", we followed a two-step process.
First, we identified holes in the topics embedding space (conceptual landscape) using persistent homology. 
Second, we used mixup barcodes to identify holes that were filled by a subset of test documents.

Persistent homology requires a filtered simplicial complex as input.
One popular semimetric to use for higher-dimensional embedding space is cosine (dis)similarity, which is particularly useful for high dimensions, where these vectors can have widely different amplitudes.
Given that our conceptual landscape is an embedding space of high-dimensional vectors, we built a filtered simplicial complex using a symmetric matrix of cosine (dis)similarity between vectors.
Using persistent homology, we identified a total of 252 one-dimensional holes in the topic embedding. 
Consider the hole represented as a red dot in \cref{pd_topic_embed_152} (a). 
This hole corresponds to a hole in the conceptual landscape shown in \cref{pd_topic_embed_152} (b). 
A hole is formed when there are topics on the perimeter that form a closed loop around negative space.
All holes, including those not visualized, contain negative space.
As a reminder, given that we are identifying holes in the embedded topics, not documents (or words), these negative spaces can also represent a low-density region in the underlying document distribution.
This means that negative spaces may contain publications that are not coherent enough to form a single topic.

To better understand the nature of the work that tends to occupy negative space, we compare research published after the training data (2021 - 2023) with documents published before the training data (2001 - 2007).
We introduce documents from these periods into the model \textit{without} retraining to accommodate the new documents and new words, as doing so would result in a re-arrangement of the embedding space.
As noted above, documents from periods of time preceding the training data should contain contextual information omitted from the training data, where documents from periods following the training data should include some amount of recombinant innovation in the form of interdisciplinary scholarship.
Our assumption is that the documents occupying these holes will combine the topics around the periphery of the hole, and that these documents will represent one or both of missing context and innovative interdisciplinarity.
To test this, we split the corpus into a training dataset (2008 - 2020) and two mutually disjoint conditions: a pre-training set (2001 - 2007) aka \textit{Class A} and a post-training set (2021 - 2023) aka \textit{Class B}.
Class A (2001 - 2007) consisted of 20,139 documents, Class B (2021 - 2023) consisted of 20,139 documents, and the training set (2008 - 2020) consisted of 73,302 documents.

Measuring the extent that Class A and Class B documents filled holes identified via persistent homology required an appropriate summary statistic.
Using persistent homology we can find the representative cycle of a hole.
This representative cycle forms the boundary of the hole (peripheral topics), as in \cref{pd_topic_embed_152}.
In 2-dimensions, the centroid will be inside a hole, provided the boundary forms a convex polygon. 
Thus, if one can find documents close to the centroid, 
in some special cases those documents will also be inside the hole.
However, if the boundary of the hole is not a convex polygon,
as will often be the case in higher dimensions,
then the centroid could not be used to find documents filling a hole.
Consequently, we opted to utilize total mixup (\cref{sec_mixup}). 
As noted above, mixup is a method specifically intended to assess the extent that new data points added to a distribution fill holes identified via persistent homology.
The total mixup summary statistic assesses the overall extent that a set of new data points, in our case documents in the Class A and Class B sets, fill the holes in the distribution.
The total mixup statistic is defined as the sum of all mixup $\frac{d-d^{'}}{d-b}$ for $1$-dimensional features.
Thus, we are able to observe the birth and death of holes in the topics embedding, and then assess the extent that additional publications drawn from Class A and B alter the death of these holes.
Documents that prompt the early death of holes in the distribution are more likely to occupy negative space \cite{Wagner:2024aa}.
A mixup simplex is the simplex which causes earlier death of a hole.
This mixup simplex contains 3 points, among these 3 points at least 1 point will be a document from the class that we added, while remaining points are topics.

Adding all the publications in the Class A and Class B sets at once and calculating total mixup is computationally expensive. 
Moreover, not all publications in a class will fill holes; some of these publications will necessarily fall outside the holes.
To reduce computational cost, we iteratively added smaller, randomly sampled subsets of documents, each consisting of 10\% of the Class A and B corpora, then calculated total mixup.
We performed 100 iterations for each of the two corpora, producing 100 measurements of total mixup for each of the two experimental conditions. 
Using permutation test \cite{pitman1937significance}, we compare the distribution of total mixups across two experimental conditions to test whether they were drawn from substantively different populations.

\begin{figure}\centering
    \subfloat[\centering Publication Histogram]{{\includegraphics[width=0.45\linewidth]{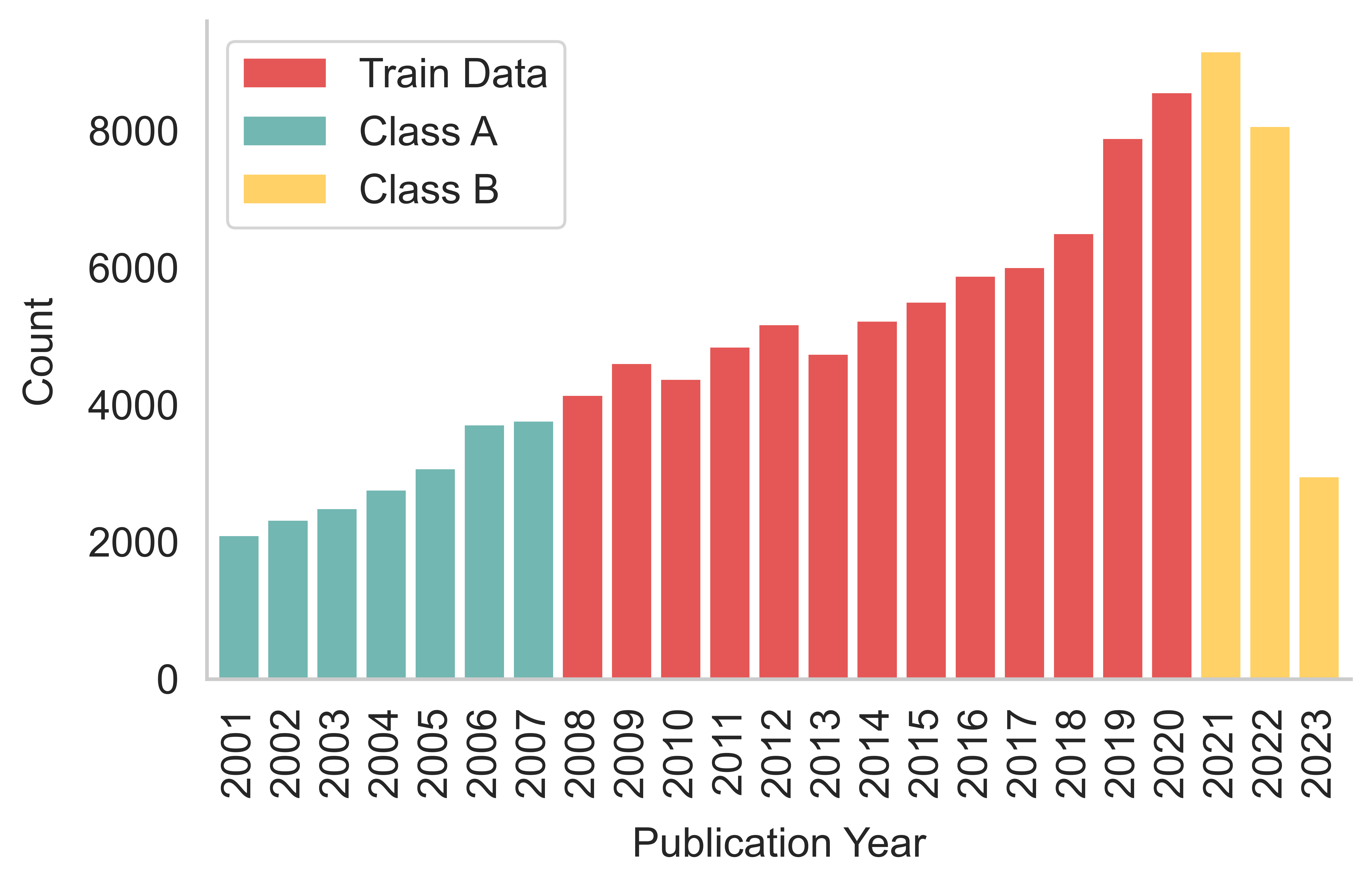} }}%
    \subfloat[\centering Histogram of total mixups]{{\includegraphics[width=0.45\linewidth]{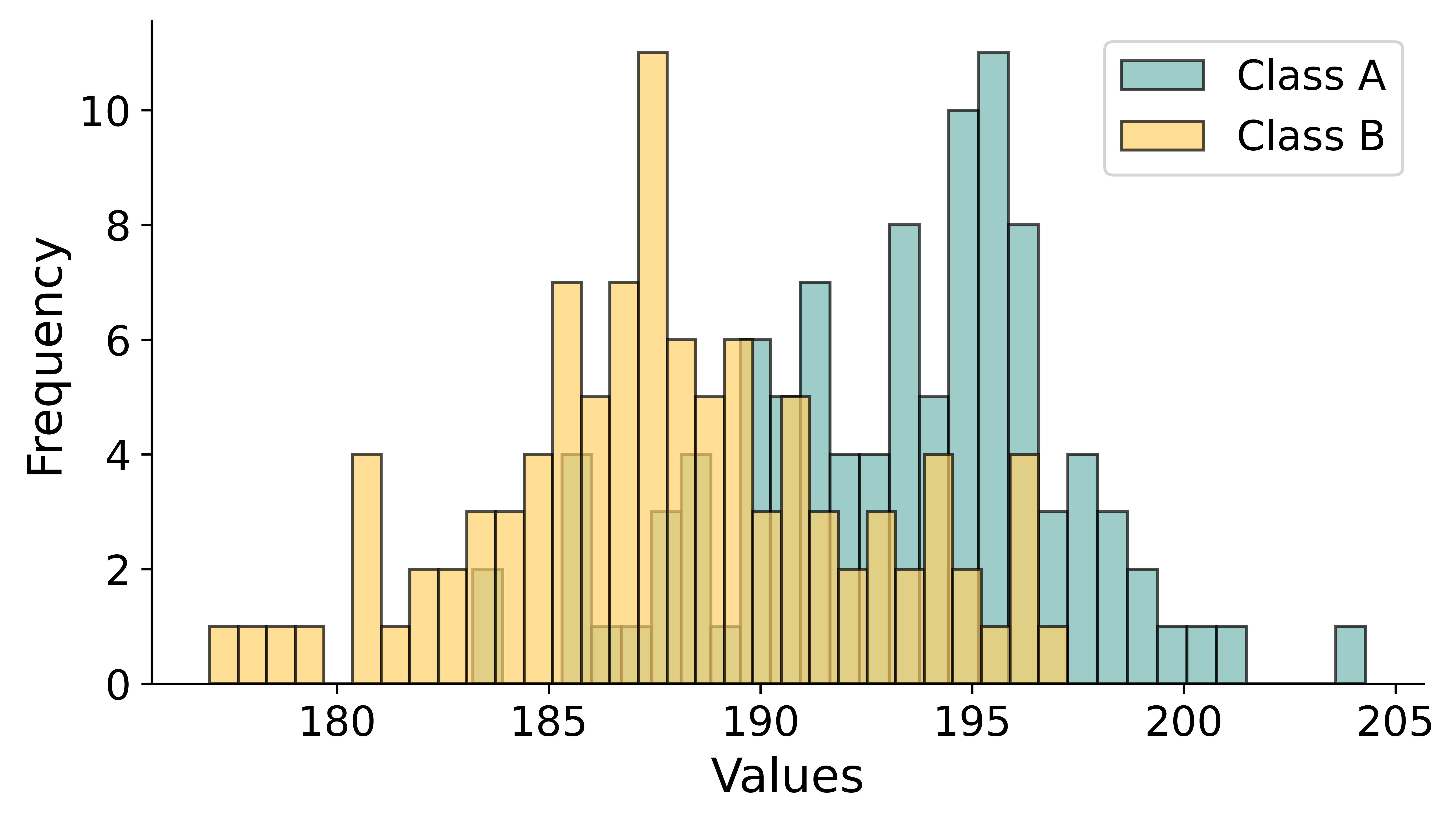} }}%
    
    \subfloat[\centering Box plot of total mixups]{{\includegraphics[width=0.45\linewidth]{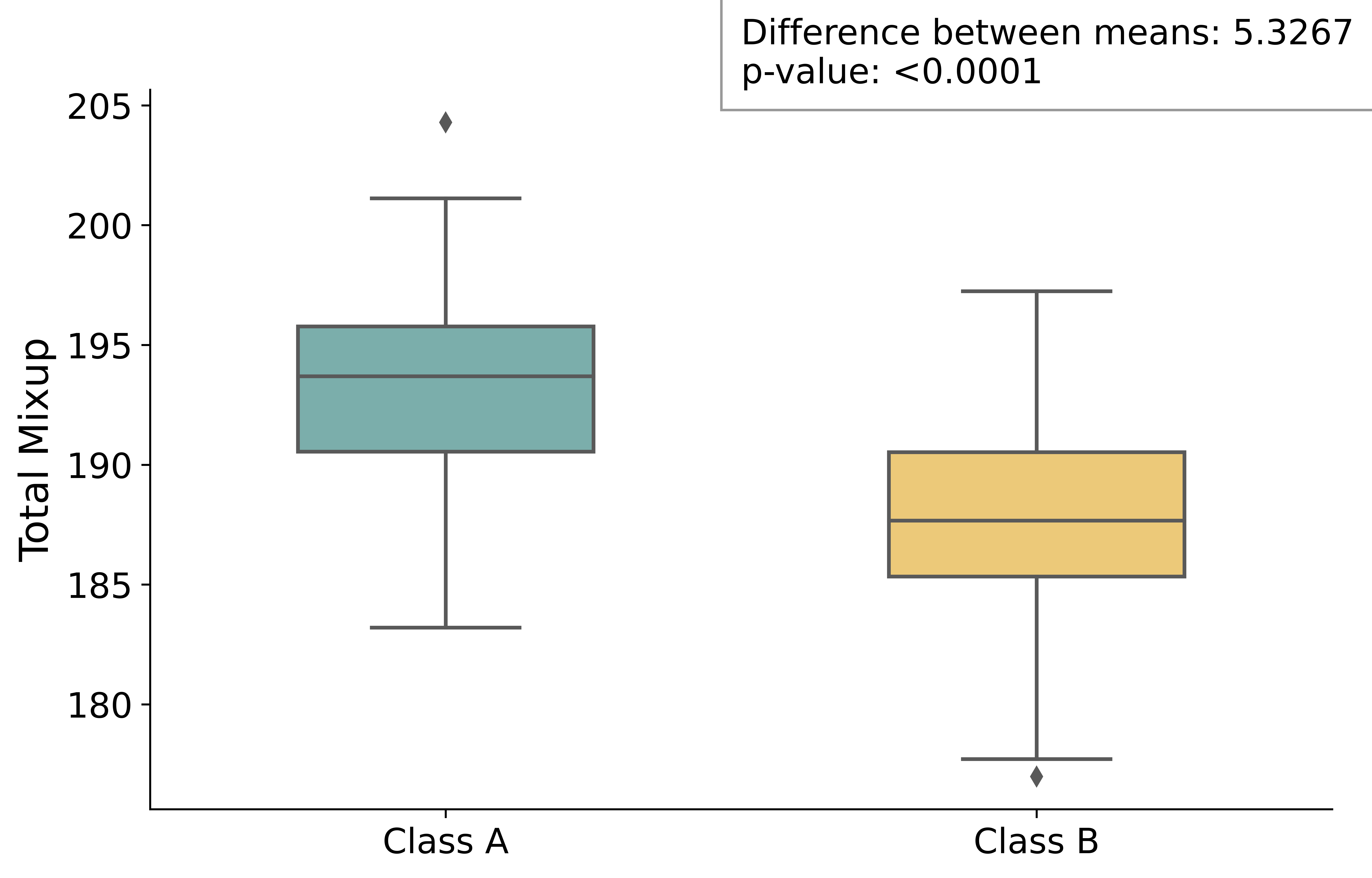} }}%
    
    \caption{
    In (a), training data are the articles which were used to generate topics. Classes A and B are the articles, some of which were added later for analysis.
    Total mixup is calculated in 100 iterations for each class.
    In one iteration, 10 percent of randomly sampled points from a class are added to topics embedding, and total mixup was recorded. 
    Histogram (b) of total mixups.
    Box plot (c) represents the distribution of total mixup for each class.
    P-value is calculated using permutation test, 
    where the observed test statistics (absolute difference between means)
    was compared with 10,000 random permutations of the dataset.
        }%
    
    \label{second_exp}%
\end{figure}

\section{Results}

In the \cref{sub:res_total_mix}, we provide results from a mean difference test comparing embedding location relative to negative space by time of publication.
This section analyzes distributions of total mixup derived of 100 subsample iterations. 
As a reminder, total mixup is a global measure of the extent that documents added into the embeddings space during mixup fill negative space.
In the \cref{sub:res_mix}, we provide a localized, descriptive analysis of specific, noteworthy holes and the documents that "filled" these holes during mixup.
In this section, we instead use mixup to identify documents filling the negative space of specific holes.

\subsection{Total mixup}\label{sub:res_total_mix}

In \cref{second_exp}, we performed analyses comparing Class A and Class B.
We observed a consistent result: Class A publications are more likely to fill holes compared to Class B publications. 
That is, publications preceding the training data are more likely to fill up the holes in topic embeddings space than publications proceeding the training data.

Average total mixup for publications in Class A is $193.2258$ and Class B is $187.8991$.
The observed difference between their average is $5.3267$.
To determine whether our observed difference is statistically significant,
we performed a permutation test.
We merged total mixups for both the classes together and then randomly permuted the data 10000 times.
Each time, we recorded the absolute difference between average total mixup and then compared it with our observed difference of $5.3267$.
There were less than $0.01$ \% cases where the absolute difference between average total mixup was higher than the observed difference, giving us a $p$-value less than $0.0001$.
This implies that Class A and Class B objects are highly likely to have consistent, statistically different total mixups.
Put simply, articles from Class A are more likely to fill up holes made by the topics compared to Class B. 

\subsection{Mixup}\label{sub:res_mix}

In the preceding subsection, we employed total mixup to capture the association between documents' time of publication and embedding relative to negative embedding space.
However, another approach to utilizing mixup is to investigate how distinct classes of documents are located relative to specific, substantive cavities in the embedding space.
There are numerous holes in the embedding space we can examine, but the most significant holes are those that persist for longer periods in the filtration (i.e., longer period of time between birth and death). 
We use a persistence value of $0.2$ to filter out intermittent cavities.
This will allow us to analyze specific effect of additional documents to individual holes.

\begin{figure}\centering
    \subfloat[\centering Persistence diagram]{{\includegraphics[width=0.5\linewidth]{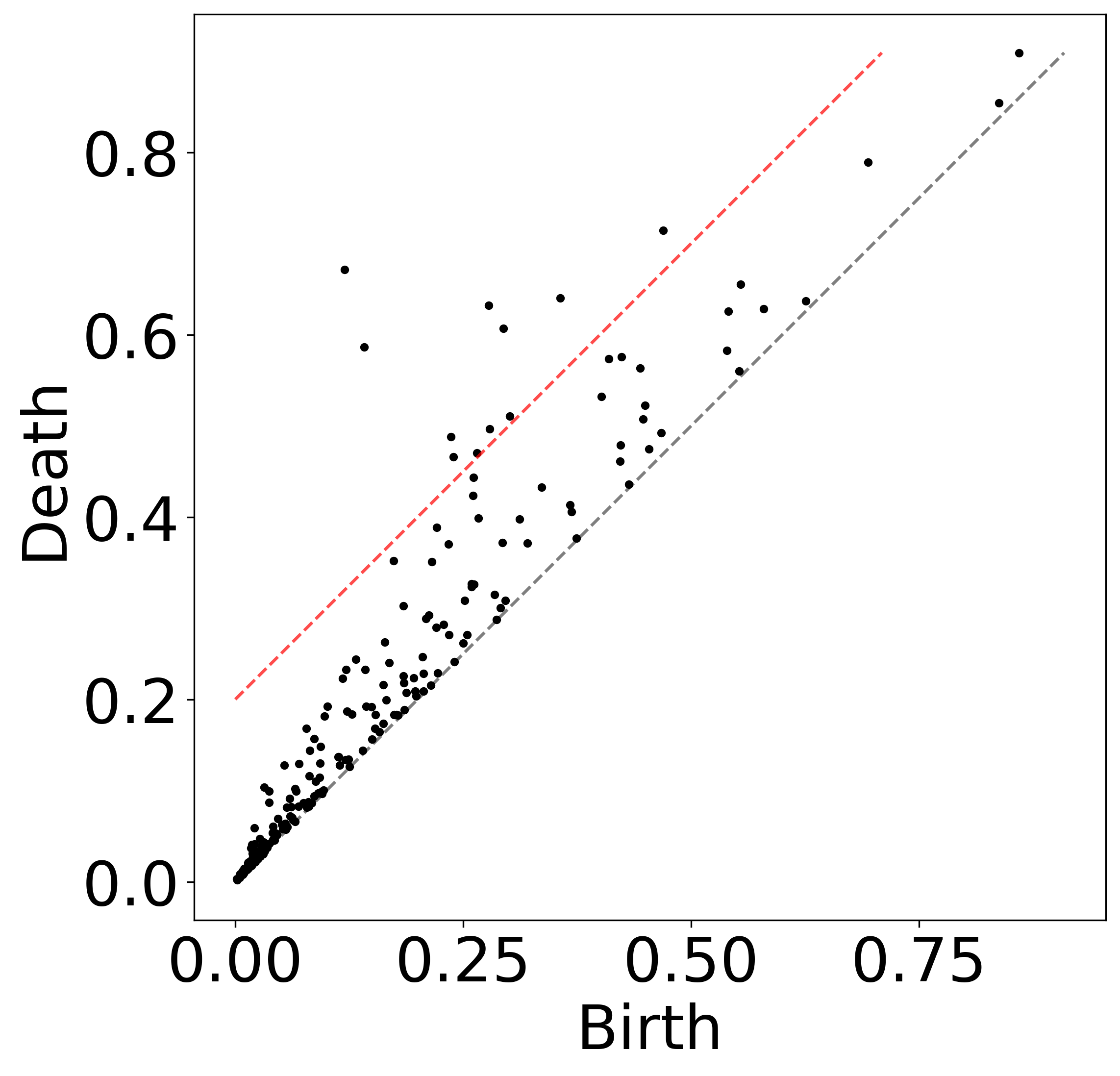} }}%
    
    \subfloat[\centering Average mixup values for different classes]{{\raisebox{1.2\height}
        {\footnotesize\label{pers_table}
        \begin{tabular}{c c c c c c}
        \hline
        \multirow{2}{*}{Index} & \multirow{2}{*}{birth} & \multirow{2}{*}{death} & \multirow{2}{*}{persistence} & \multicolumn{2}{c}{Average mixup} \\
        \cline{5-6}
        & & & & Class A & Class B \\
        \hline
        1  & 0.119558 & 0.671046 & 0.551488 & 0.123763 & 0.146369 \\
        2  & 0.141169 & 0.586396 & 0.445228 & 0.110025 & 0.106691 \\
        3  & 0.277931 & 0.632140 & 0.354209 & 0.111869 & 0.047935 \\
        4  & 0.294109 & 0.606568 & 0.312459 & 0.175501 & 0.104154 \\
        5  & 0.356116 & 0.640193 & 0.284076 & 0.239535 & 0.242829 \\
        6  & 0.236416 & 0.488102 & 0.251687 & 0.301283 & 0.238126 \\
        7  & 0.469428 & 0.714105 & 0.244677 & 0.969302 & 1.000000 \\
        8  & 0.239154 & 0.465921 & 0.226767 & 0.425958 & 0.183792 \\
        9  & 0.278753 & 0.496464 & 0.217710 & 0.382394 & 0.507305 \\
        10 & 0.300954 & 0.510708 & 0.209755 & 0.190748 & 0.070388 \\
        11 & 0.265018 & 0.470292 & 0.205274 & 0.767211 & 0.596149 \\
        \hline
        \end{tabular}
        }
    }}%
    
    \caption{
    (a) Points above the dotted red line has persistence greater than $0.2$.
    There are 11 such points above the dotted red line.
    (b) Average mixup information for these 11 points when documents from classes A and B are added into the embedding space.
        }%
    
    \label{pd_threshold}%
\end{figure}

In \cref{pd_threshold}, there are eleven points on the persistence diagram with persistence greater than $0.2$, representing eleven persistent holes in the topic embeddings space.
Contrary to our previous analysis of the total mixup, which observes a consistent trend where total mixup for Class A documents is more likely to be significantly greater than the total mixup for Class B documents, we observed greater mixup for Class B than Class A in four of these persistent holes.
This is an interesting observation, as it indicates that a sizable proportion of persistent holes in the topic embeddings space are being filled more by documents published after the training data (rather than before). 

\subsubsection{Documents permeating the ninth representative cycle}

\begin{figure}\centering
    \subfloat[\centering Histogram]{{\includegraphics[width=0.4\linewidth]{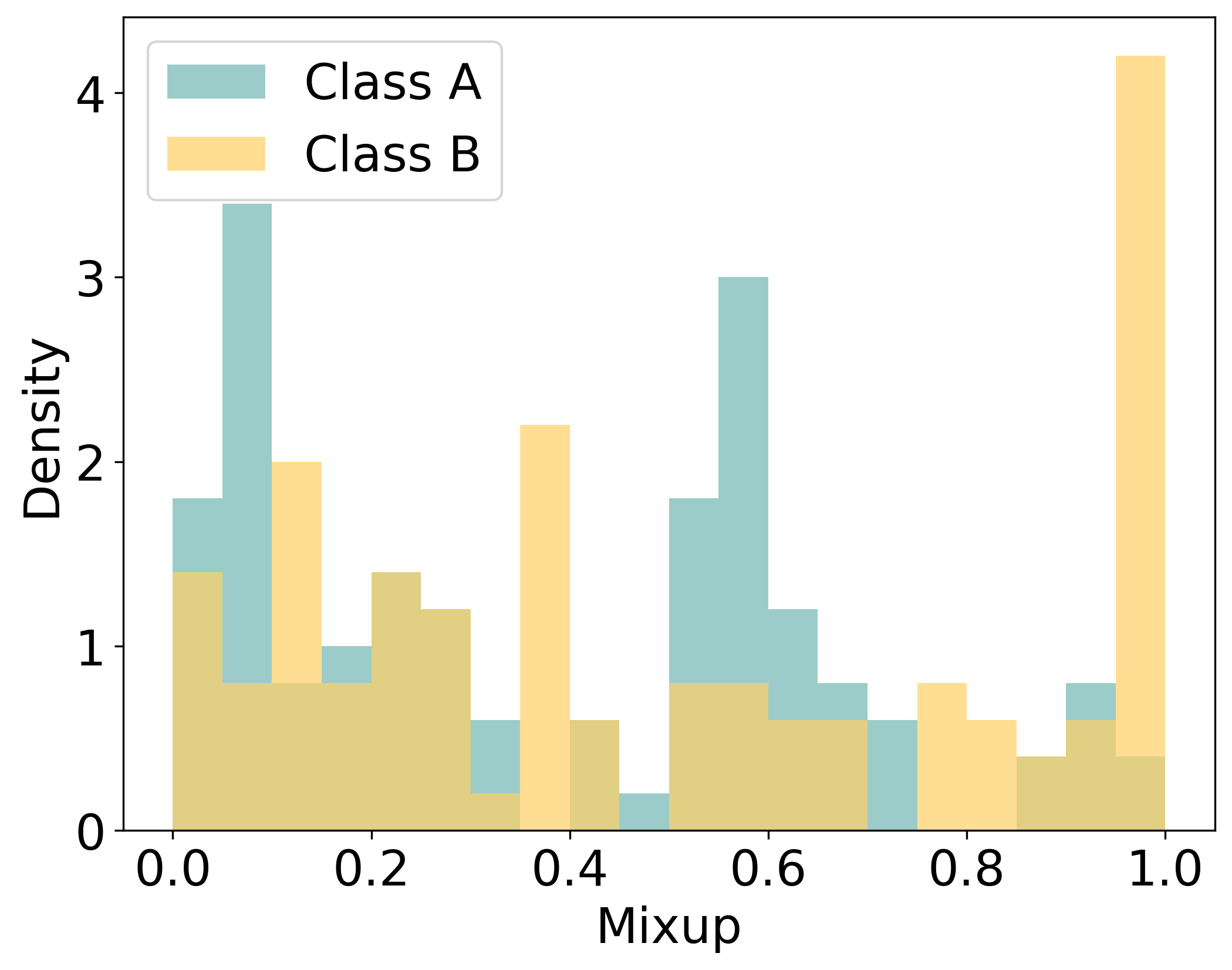} }}%
    \subfloat[\centering Box plot]{{\includegraphics[width=0.4\linewidth]{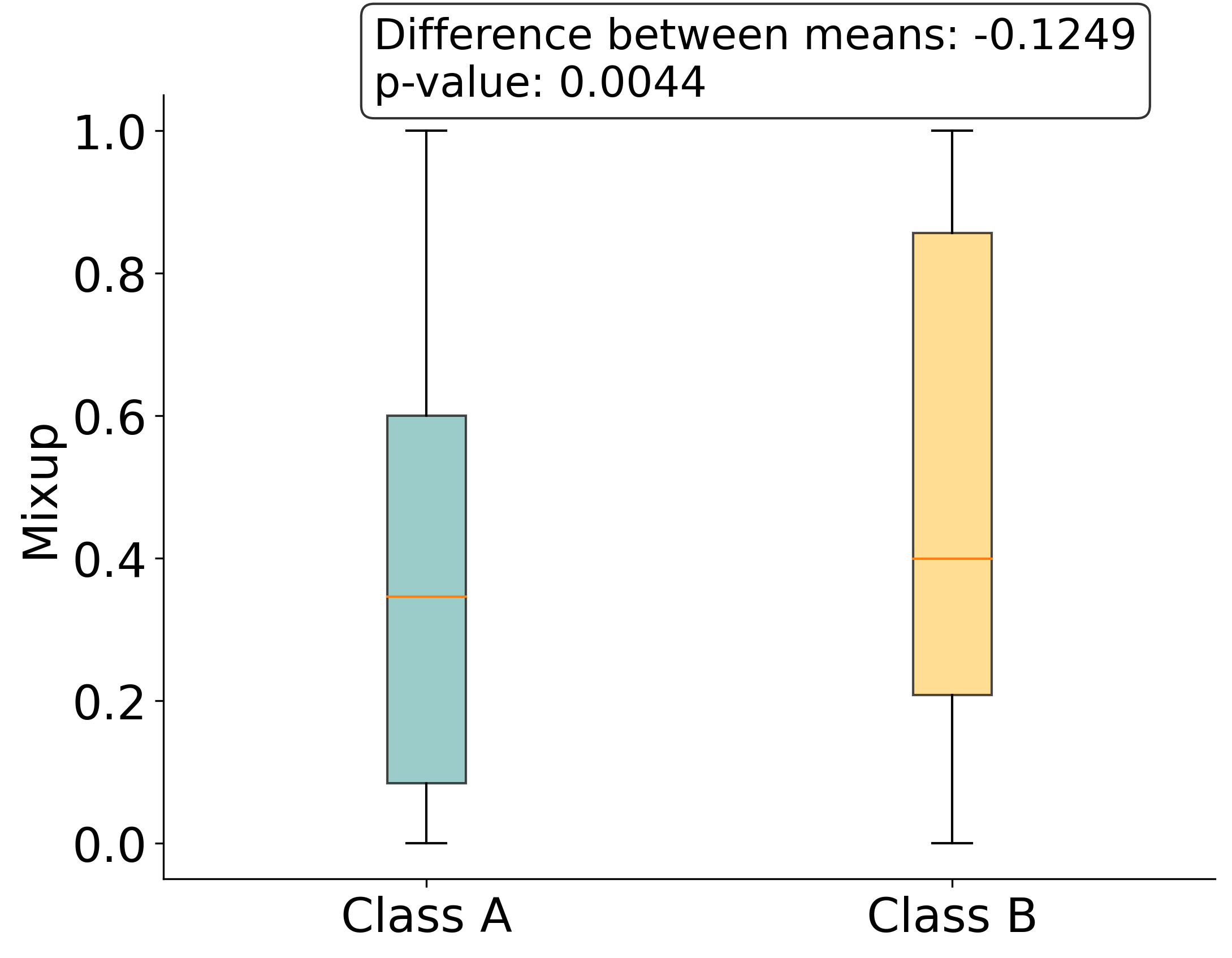} }}%

    \caption{
    (a) Histogram for mixup distribution for holes indexed by $9$.
    (b) Box plot of the two mixup distribution for class A and B visualizing the statistical difference between them using observed difference between means and applying permutation test for $10000$ iterations.
        }%
    \label{box_hist_243}%
\end{figure}

In \cref{pers_table}, the index 9 hole in the table exhibits a higher average mixup for Class B compared to Class A.
This implies that documents belonging to Class B are on average more likely to occupy the hole than those belonging to Class A.
This observation is further strengthened by the comparison of the underlying distribution of mixup values.
In \cref{box_hist_243}, we compared the mixup distributions for both classes A and B mixup documents and observed a statistically significant difference.

Interpreting the topics included in the representative cycle of this hole, as well as some of the documents that fill the hole, it is much easier to understand this phenomenon. 
In total, there are 12 topics in the representative cycles.
These topics can be further classified into five subject classifications: life sciences \& biology,
physical sciences \& engineering, 
agricultural \& environmental
and
medicine \& healthcare.
Each classification contains 6, 3, 2 and 1 topics, respectively, from the representative cycle.

\begin{figure}\centering
    \subfloat[\centering]{{\includegraphics[width=0.4\linewidth]{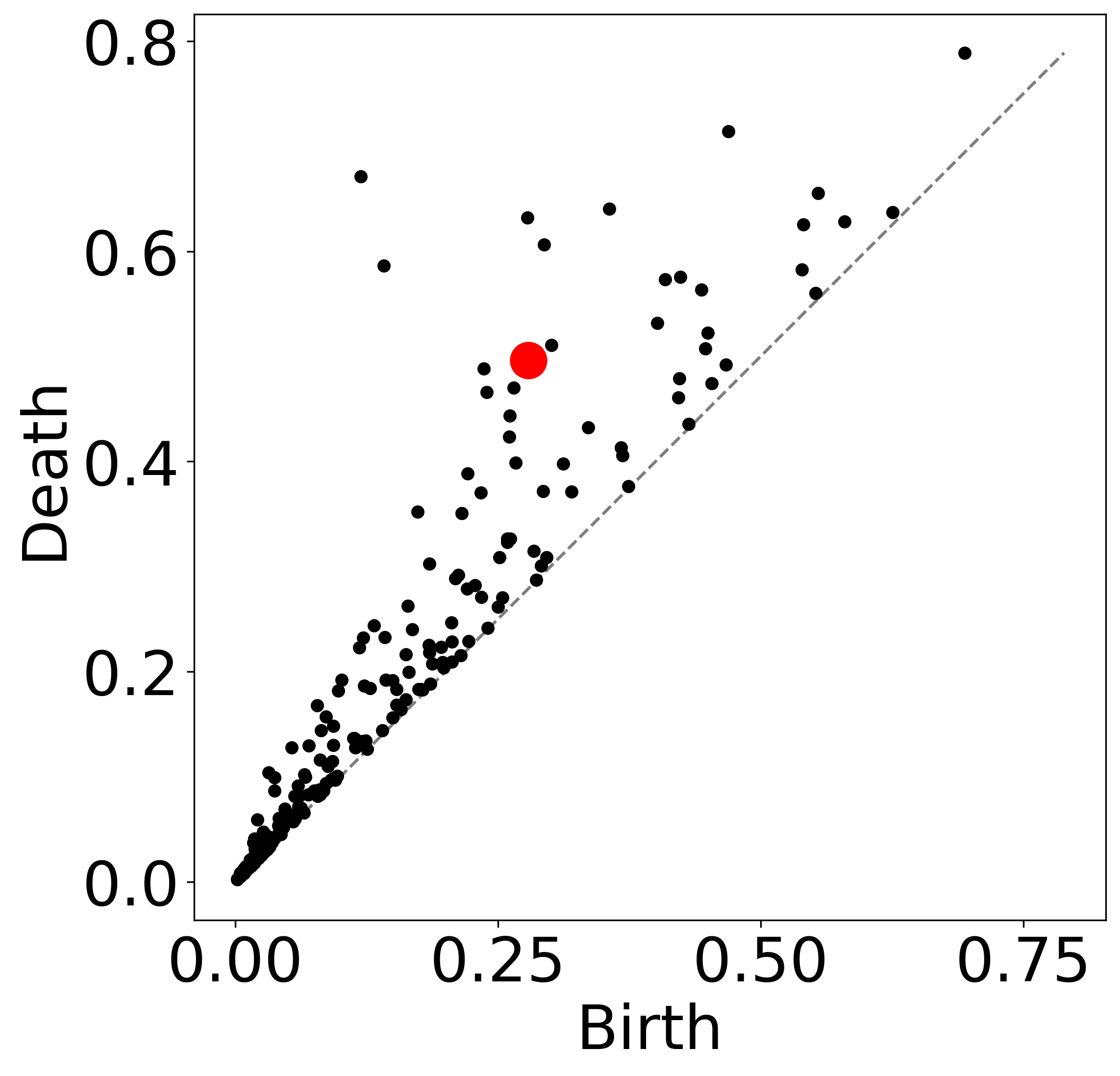} }}%
    
    \subfloat[\centering]{{\includegraphics[width=0.4\linewidth]{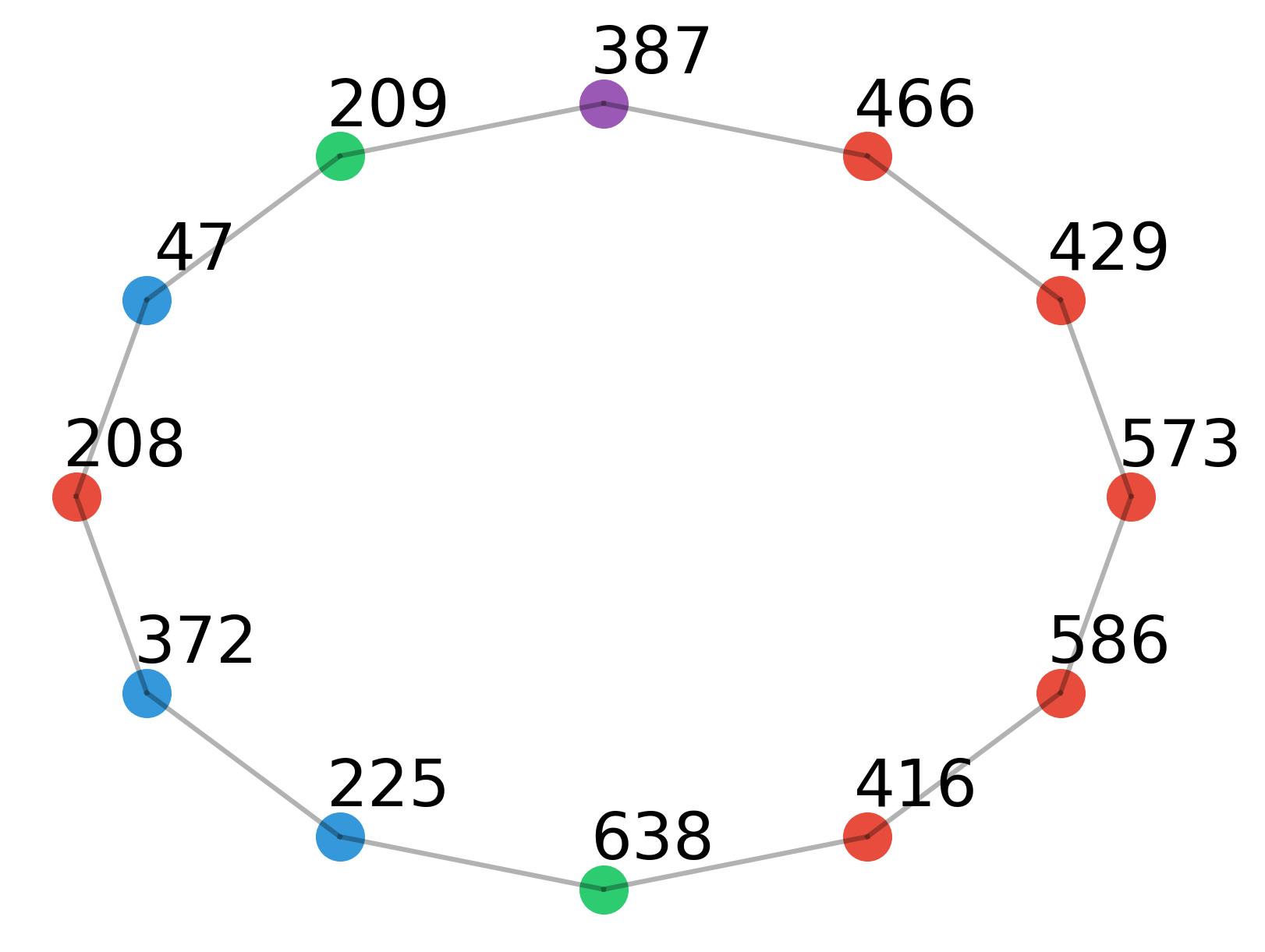} }}%
    \subfloat[\centering]{{\includegraphics[width=0.4\linewidth]{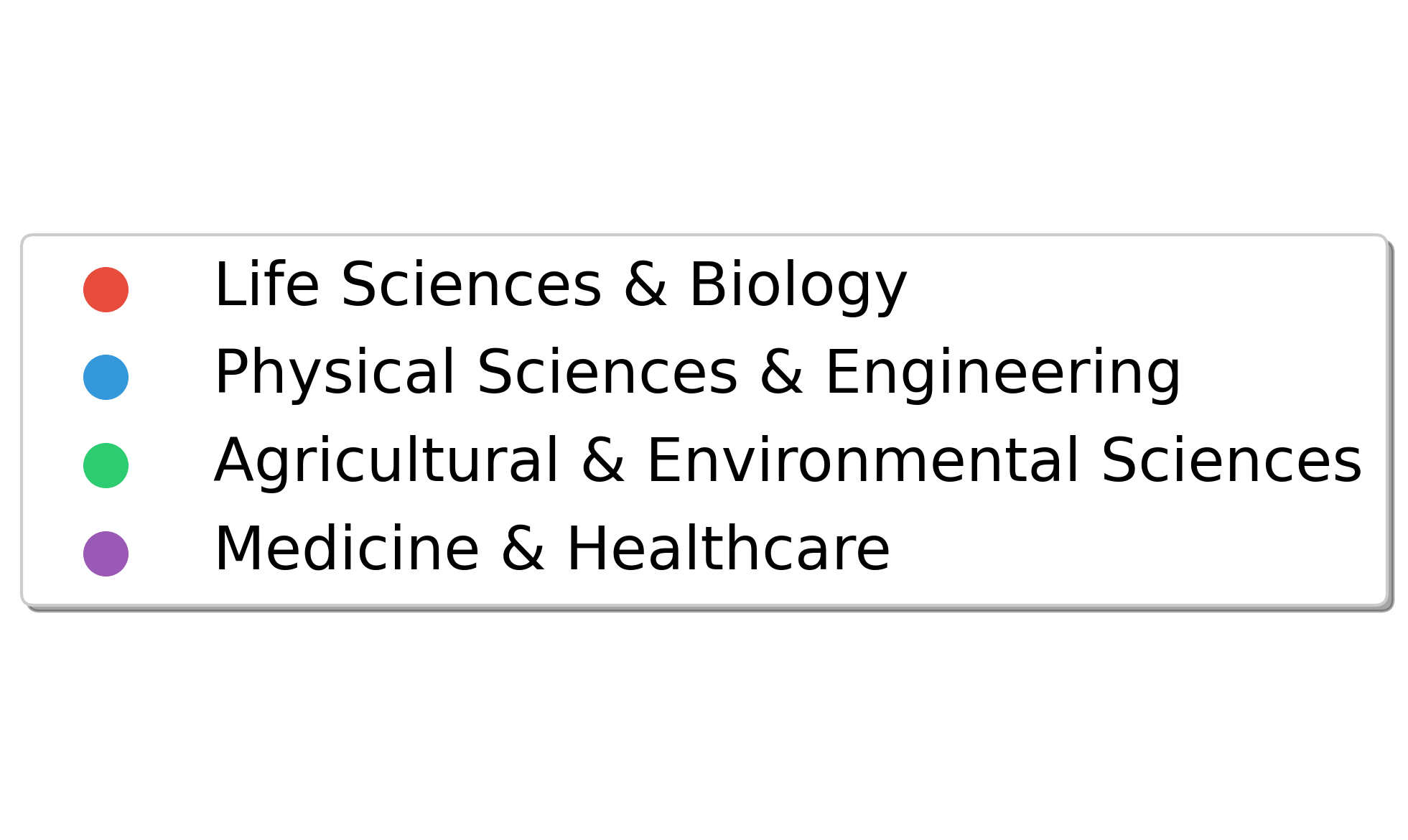} }}%

    \subfloat[\centering]{{\includegraphics[width=0.4\linewidth]{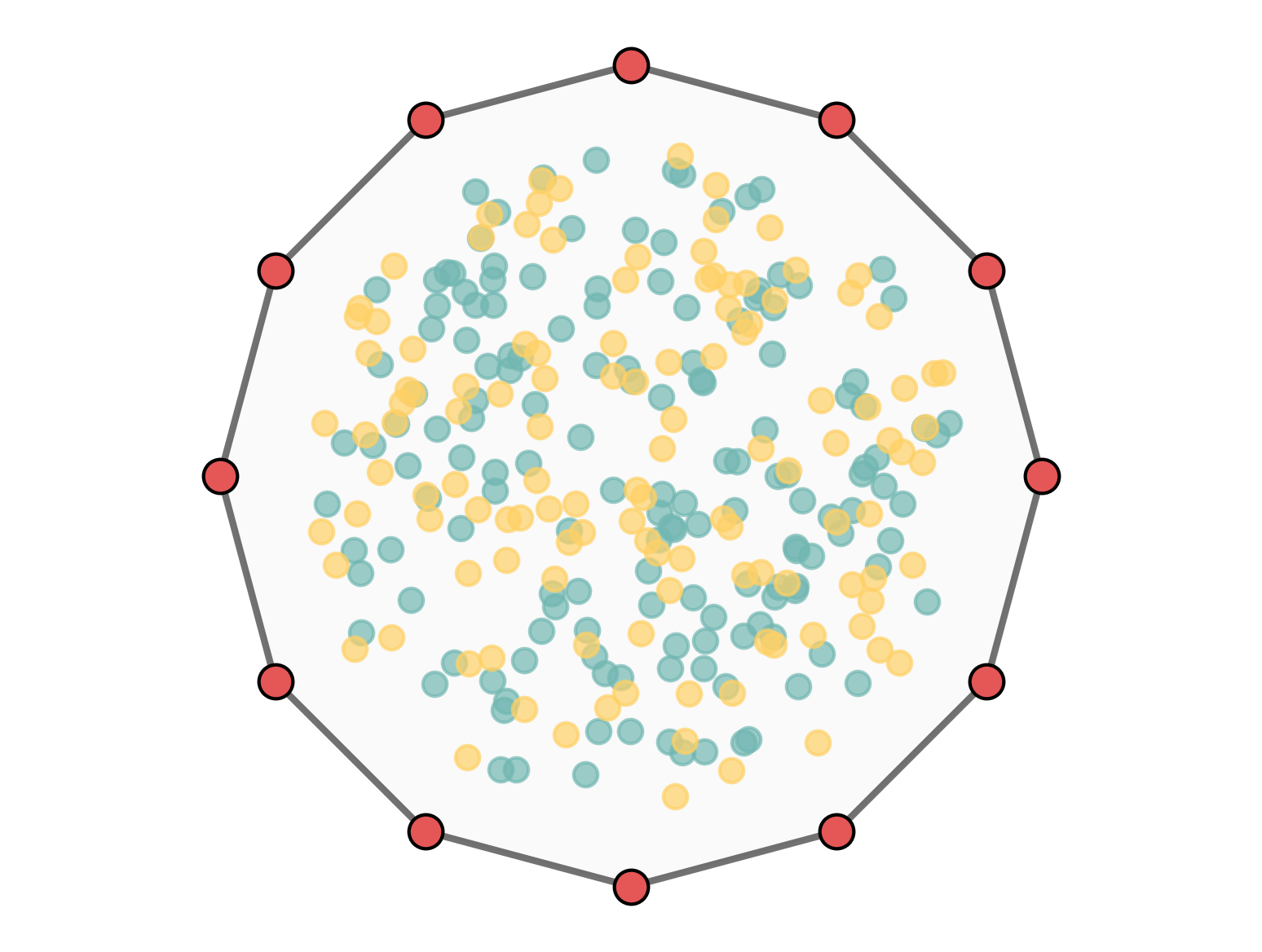} }}%
    \subfloat[\centering]{{\includegraphics[width=0.4\linewidth]{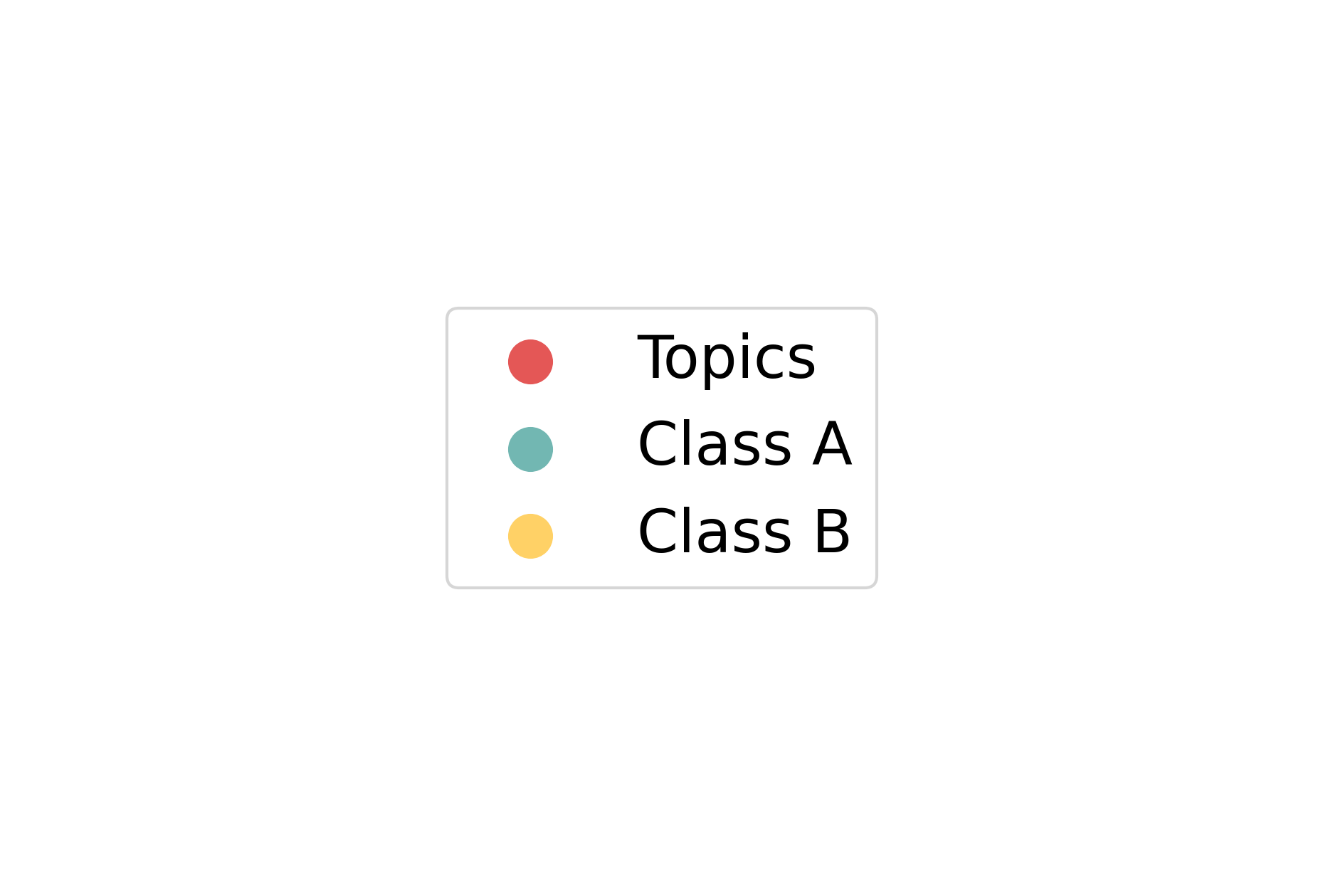} }}%
    
    \caption{(a) Persistence diagram with index 9 hole.
    Since our data is in $5$ dimensions, we use a polygon representation
    for visualization.
    (b) and (c) Topics labeled with their corresponding subject classification.
    (d) and (e) Representation of Classes A and B documents filling the negative space made by index 9 hole.
    }%
    \label{fill_243}%
\end{figure}

During our analysis of incorporating mixup document classes A and B into the embedding space (refer to \cref{fill_243}), the negative space of the focal hole was populated with 154 and 123 documents belonging to classes A and B, respectively.
Although Class A contained 31 more documents within the hole, the average mixup was lower compared to Class B.
This observation underscores a key advantage of using mixup techniques when assessing the suffusion of negative space, rather than relying on the count of points that are located within that space. 
This is, of course, in addition to the aforementioned benefit wherein mixup does not require convex representative cycles, as would be the case for assessing proximity to a centroid.

The Class B mixup documents that fill the focal hole span multiple established topic areas. The titles of a small selection of documents, as well as short descriptions of their interdisciplinary themes, are reported in \cref{tab:243}. 

\begin{small}
\begin{longtable}{>{\raggedright}p{0.25\textwidth}>{\raggedright}p{0.3\textwidth}>{\raggedright\arraybackslash}p{0.35\textwidth}}
    \textbf{Interdisciplinary Area} & \textbf{Research Title} & \textbf{Description} \\
    \hline
    \endhead
    Transplant Medicine + Vascular Biology & ``A novel injury site‐natural antibody targeted complement inhibitor protects against lung transplant injury''\cite{li2021novel} & Combines transplant medicine with vascular biology and complement system research \\
    \hline
    Transplant Medicine + Cellular Biology & ``Resolution of post-lung transplant ischemia-reperfusion injury is modulated via Resolvin D1-FPR2 and Maresin 1-LGR6 signaling''\cite{leroy2023resolution} & Bridges transplant surgery with cellular stress response and signaling mechanisms \\
    \hline
    Pediatric Oncology + Pharmacogenomics & ``SAMHD1 Single Nucleotide Polymorphisms Impact Outcome in Children with Newly Diagnosed Acute Myeloid Leukemia''\cite{marrero2023samhd1} & Combines pediatric cancer treatment with genetic biomarker analysis \\
    \hline
    Pediatric Oncology + Pharmacogenomics + Precision Medicine & ``Polygenic Ara-C Response Score Identifies Pediatric Patients With Acute Myeloid Leukemia in Need of Chemotherapy Augmentation''\cite{elsayed2022polygenic} & Bridges oncology, pharmacogenomics, and precision medicine approaches \\
    \hline
    Pulmonary Medicine + Molecular Biology + Targeted Therapy & ``Lipoxin A4 mitigates ferroptosis via FPR2 signaling during lung ischemia-reperfusion injury''\cite{sharma2022lipoxin} & Combines pulmonary medicine, cellular death mechanisms, and targeted therapeutics \\
    \hline
    \caption{
    Table describing different documents from class B which were filling up the hole index by $9$. 
    These are interdisciplinary research combining multiple fields including transplant medicine, 
    pediatric oncology, critical care, and pulmonary medicine with various biological sciences.}
    \label{tab:243}
\end{longtable}
\end{small}

\subsubsection{Documents permeating the eleventh representative cycle}

\begin{figure}\centering
    \subfloat[\centering Histogram]{{\includegraphics[width=0.4\linewidth]{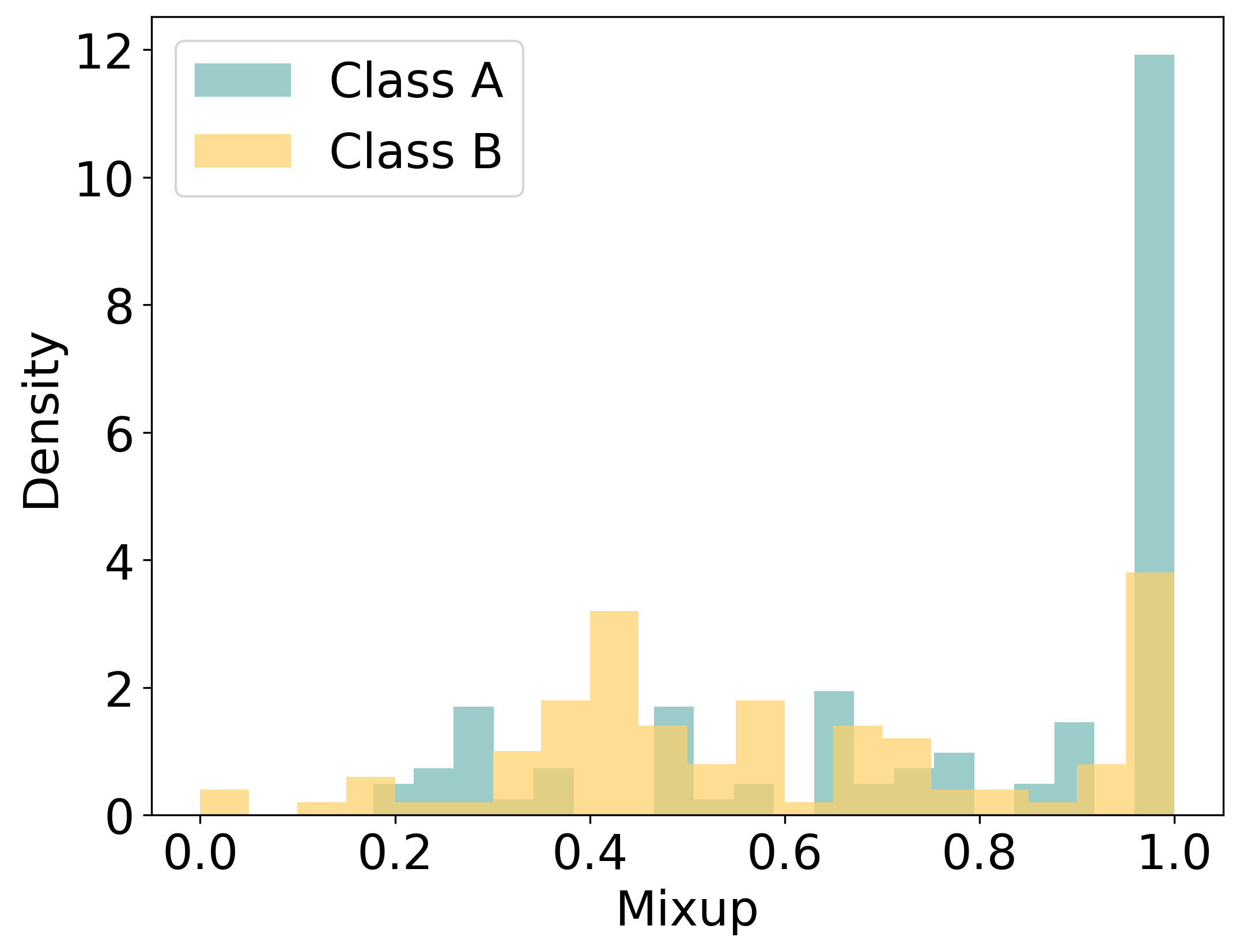} }}%
    \subfloat[\centering Box plot]{{\includegraphics[width=0.4\linewidth]{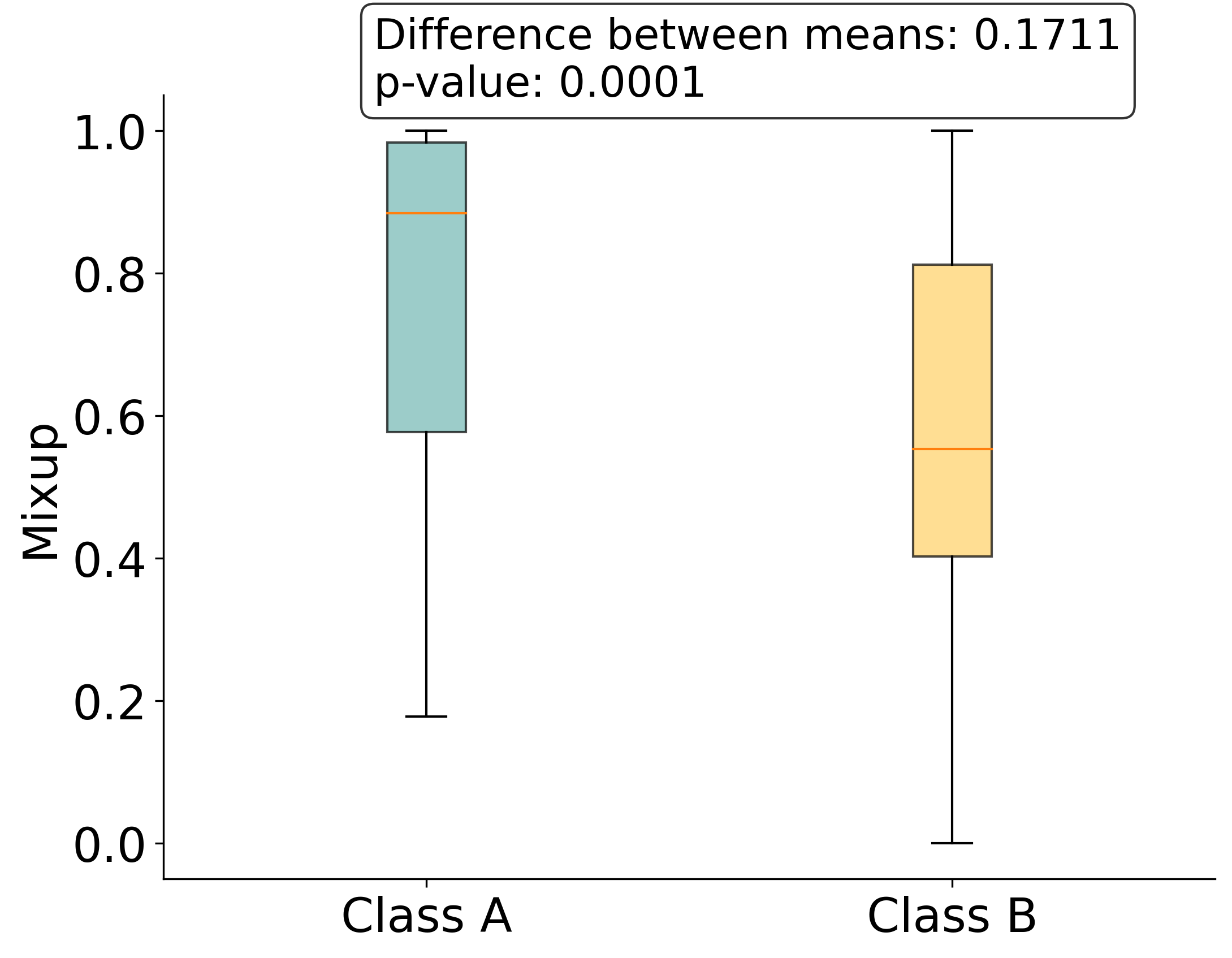} }}%

    \caption{
    (a) Histogram for mixup distribution for holes indexed by $11$.
    (b) Box plot of the two mixup distribution for class A and B visualizing the statistical difference between them using observed difference between means and applying permutation test for $10000$ iterations
        }%
    
    \label{box_hist_241}%
\end{figure}

The eleventh hole reported in \cref{pers_table} exhibits a higher average mixup for Class A compared to Class B.
This implies that documents published before the training set are, on average, more likely to occupy this hole than those published after the training set.
As before, we compared the mixup distributions for both classes A and B documents and observed a significant statistical difference between them (see \cref{box_hist_241}.)
In total, there were 10 topics in the representative cycle.
These topics were classified into five subjects:
medicine \& healthcare, 
physical sciences \& engineering, 
life sciences \& biology,
social science
and agricultural extension.
Each classification contains 4, 3, 1, 1 and 1 topics, respectively, from the representative cycle.

\begin{figure}\centering
    \subfloat[\centering]{{\includegraphics[width=0.4\linewidth]{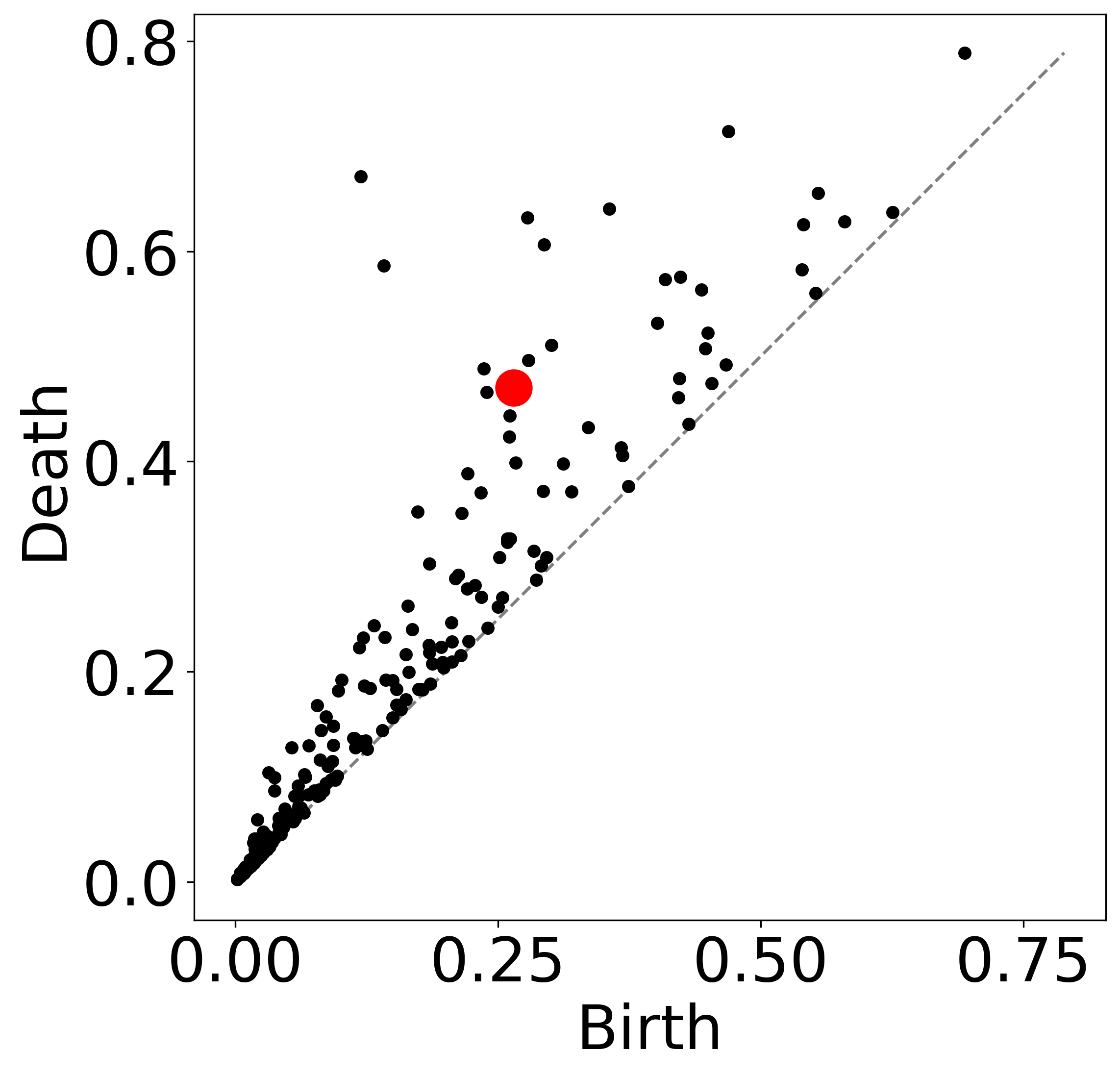} }}%
    
    \subfloat[\centering]{{\includegraphics[width=0.4\linewidth]{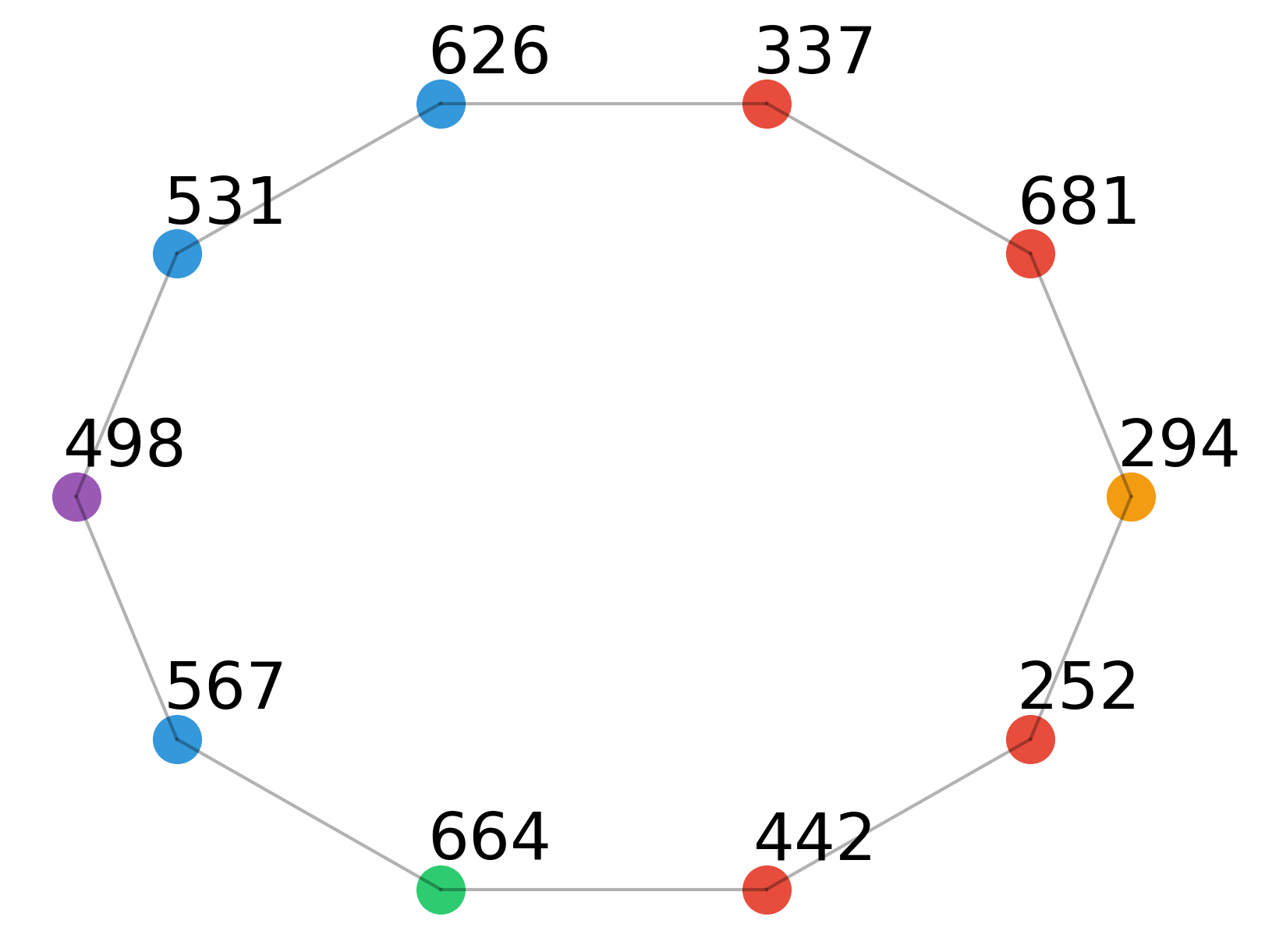} }}%
    \subfloat[\centering]{{\includegraphics[width=0.4\linewidth]{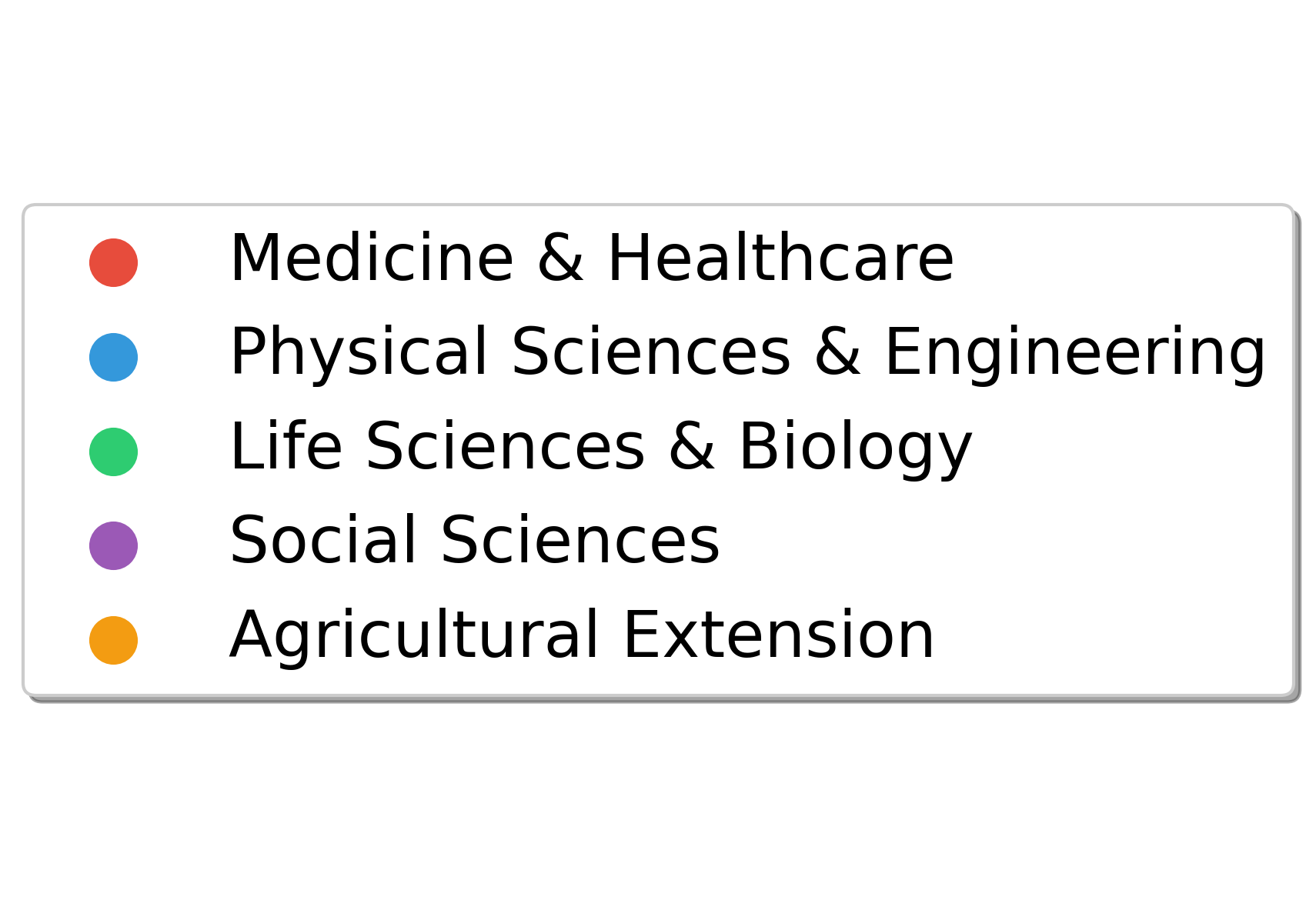} }}%

    \subfloat[\centering]{{\includegraphics[width=0.4\linewidth]{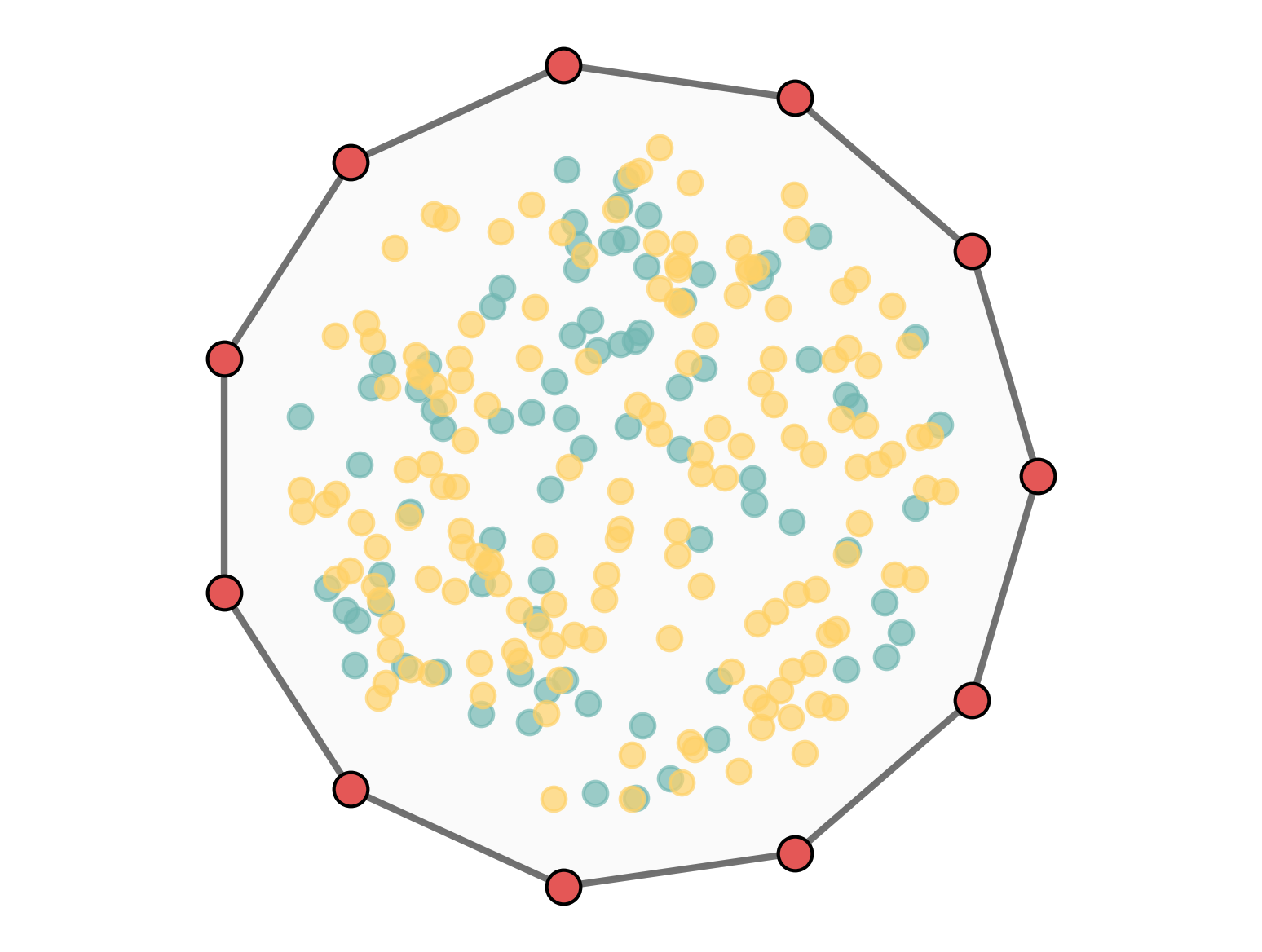} }}%
    \subfloat[\centering]{{\includegraphics[width=0.4\linewidth]{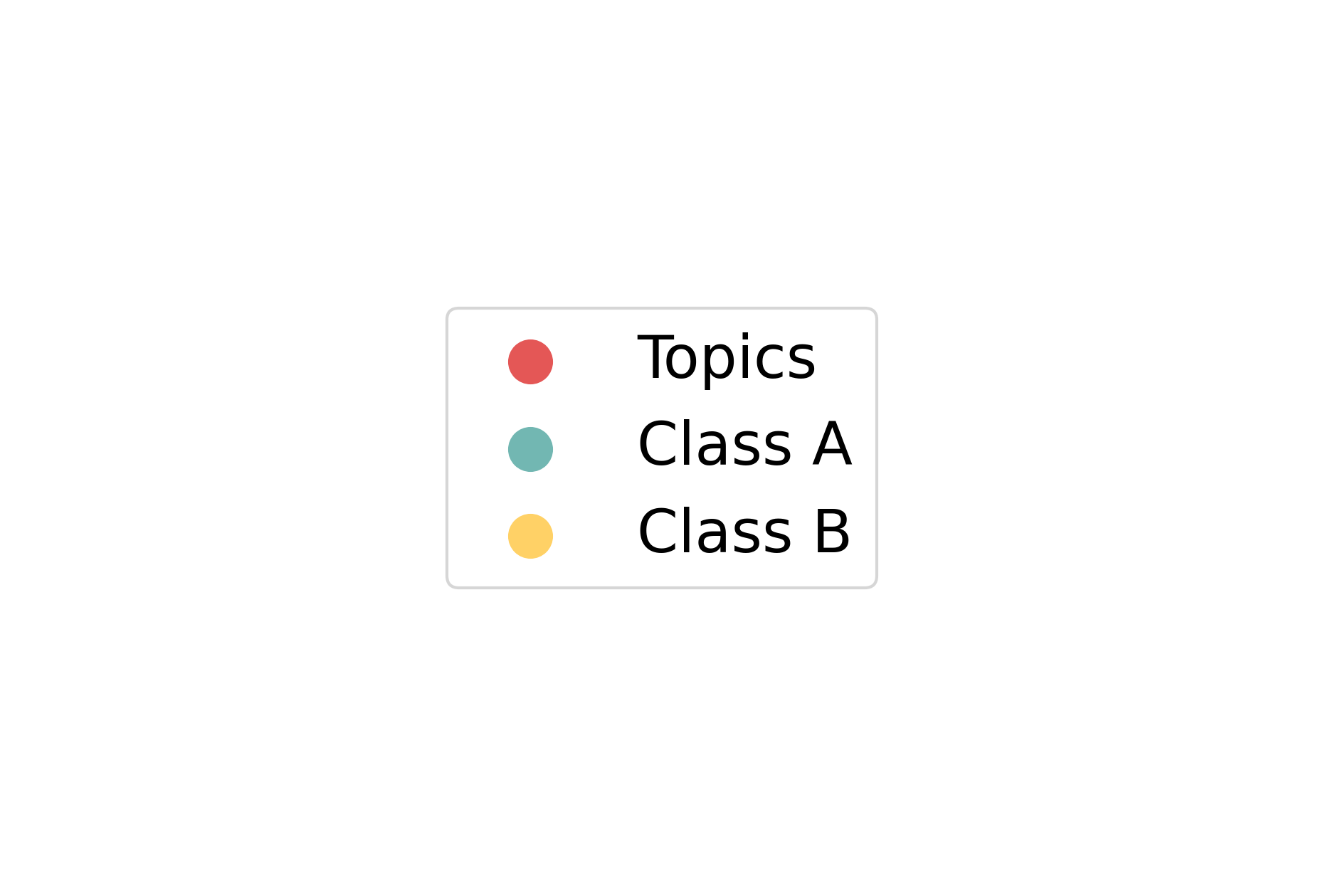} }}%
    
    \caption{(a) Persistence diagram with index $11$ hole.
    Since our data is in $5$ dimensions, we use a polygon representation
    for visualization.
    (b) and (c) Topics labeled with their corresponding subject classification.
    (d) and (e) Representation of Classes A and B documents filling the negative space made by index $11$ hole.
    }%
    \label{fill_241}%
\end{figure}

During our analysis of the mixup documents introduced into the embedding space (see \cref{fill_241}), we found that the negative space of the focal hole was populated by 81 mixup documents from Class A, and 155 documents from Class B.
The interdisciplinary research documents filling the focal hole in Class A include established topic areas. Examples of these documents, categorized by interdisciplinary themes, are provided in \cref{tab:241}.

\begin{small}
\begin{longtable}{>{\raggedright}p{0.25\textwidth}>{\raggedright}p{0.3\textwidth}>{\raggedright\arraybackslash}p{0.35\textwidth}}
    
    \textbf{Interdisciplinary Area} & \textbf{Research Title} & \textbf{Description} \\
    \hline
    \endhead
    Neuroscience + Metabolism + Cell Signaling & ``Chibby Promotes Adipocyte Differentiation through Inhibition of $\beta$-Catenin Signaling''\cite{li2007chibby} & Bridges Cellular/Molecular Biology with Metabolism/Appetite Regulation and developmental signaling pathways \\
    \hline
    Gene Therapy + Neuroscience + Materials Science & ``Polyethylenimine-mediated NGF gene delivery protects transected septal cholinergic neurons''\cite{wu2004polyethylenimine} & Bridges Materials Science with Neurology and Cellular Biology \\
    \hline
    Endocrinology + Neuroscience + Psychology & ``Estrogen effects on cognition and hippocampal transcription in middle-aged mice''\cite{aenlle2009estrogen} & Combines Psychology/Personality Research with Neurology and endocrinology \\
    \hline
    Stem Cell Biology + Metabolism + Developmental Biology & ``Derivation of Adipocytes from Human Embryonic Stem Cells''\cite{xiong2005derivation} & Combines Cellular Biology with Metabolism and developmental biology \\
    \hline
    Circadian Biology + Neuroscience + Cell Signaling & ``A computational model for functional mapping of genes that regulate intra-cellular circadian rhythms''\cite{liu2007computational} & Combines Cellular Biology with Neurology and computational modeling \\
    \hline
    \caption{
    Table describing different documents from class A which were filling up the hole index by $11$.
    These documents represent highly specialized molecular neuroscience research that integrates multiple biological systems,
    creating interdisciplinary connections across neuroscience, cell biology, materials science, and metabolism.}
    \label{tab:241}
\end{longtable}
\end{small}

\section{Discussion and Conclusions}

Informed by literature in science policy and innovation studies and structural linguistics, this paper presented a prospective approach to identifying and measuring the interdisciplinarity of scientific articles based on their content (or the content of their abstracts). 
Findings imply a complex relationship between the mixup scores of documents and their time of publication. 
Mixup scores are ascertained based on their placement respective to negative space (i.e., `holes') when introduced into an embedding distribution learned by a topic model \cite{Wagner:2024aa}. 
Given that we observe evidence that publications located in the negative space betwixt topics tended to integrate those topics, and are dependent on the time of publication with respect to the training set, we conclude that negative embedding space is meaningful and merits further study. 
Diverse fields of research can sometimes struggle to identify points of integration among divergent subfields \cite{Smith_Vacca_Mantegazza_Capua_2021}. 
The application of TDA and mixup to embedded scientific publication data represents an opportunity to better understand the conceptual landscape of a given field or collection of fields. 

Our initial assumption was that interdisciplinary research that bridges different topics would be more likely to occupy the negative space between those topics. 
It followed that more recent publications would be more likely to integrate multiple subfields when establishing new ideas, per the theory of \textit{recombinant innovation} \cite{Weitzman1998,FlemingEtAl2007,Leahey&Moody2014}. 
However, a more complex picture emerged.
Publications following the training data occupied negative embedding space less frequently than publications preceding the training data. 
It appears likely that the negative embedding space may represent what we have referred to as the `missing context' of publications that precede the training data.
This lends further support to research finding that unsupervised embeddings capture latent information that has not been directly observed by the model \cite{Tshitoyan_et_al_2019}; that is, historical information, not hints of future discovery.
This observation can be better explained by returning to the jigsaw puzzle analogy.
Given that unobserved publications that precede the training set occupy negative embedding space more frequently, the `holes' in the jigsaw puzzle are, quite literally, missing pieces of the puzzle that are inferred by the language model based on the puzzle pieces it can access (the training set).
On the other hand, innovative publications that follow the training set are new puzzle pieces that do not "fit" the current configuration of the puzzle.
That is to say, they contain information that the embedding model does not know how to appraise and therefore ignores (recall that \textit{top2vec} ignores words that it has not learned). 
The jigsaw puzzle would need to be disassembled and rearranged to "fit" these new pieces, meaning that the model would need to be re-trained to accommodate truly innovative, unobserved information.

Granted, this analogy is imperfect given that we needed to operate at the topic-level in order to make the analysis computationally feasible. 
As such, the negative space in distributed topics might represent low-density regions of the underlying documents, rather than an absence of documents.
When applied in this way, our approach still runs the risk of masking the influence of subfields that publish at a reduced rate, thereby generating fewer document clusters that rise to the level of becoming a topic \cite{Parish_et_al_2018}.
However, given that the default hyperparameter specification of \textit{top2vec} only requires clusters of at least 15 papers to recognize a coherent topic \cite{angelov2020}, it is unlikely this approach would disfavor humanities and social scientific scholarship to the same degree as traditional metrics (impact factor, H-index, etc.) 

In this initial work, we opted to employ the \textit{top2vec} topic embedding model, a popular model at the time the project was conceived. 
Since conception, \textit{BERTopic} has developed as an extension of \textit{top2vec}, utilizing bidirectional encoder representations from transformers (BERT), rather than the shallow neural networks from which \textit{top2vec} inherits its name \cite{grootendorst2022}. 
\textit{BERTopic} includes a host of features that would support the further testing and implementation of mixup as a means of identifying interdisciplinary research. 
Primarily, the capacity of BERT to continue pre-training from a `checkpoint' would allow future research to pre-train the underlying topic model on a larger corpus, and then further pre-train (or `fine-tune') on a training corpus, like the one presented in this paper.
In theory, the additional context offered by the larger corpus would ensure that the structure of the underlying embeddings space better represents a generalized (structural) understanding of science. 
Moreover, fewer words would be omitted since words that do not appear in the training set (e.g., University of Florida publications) could appear in the pre-training set (e.g., Wikipedia, Web of Science, or dimensions.ai). 
Put simply, pre-training on a larger corpus (e.g., Wikipedia, Web of Science, or dimensions.ai) would be similar to providing the embedding model with a reference photo of the completed puzzle, including information on holes that exist in the conceptual space of the overarching, comprehensive knowledge base.
Given our contention that the negative space in embedded documents represents missing context that a model infers based on the content of later publications, pre-training on a more exhaustive corpus and fine-tuning on the training set may produce results more in line with our initial assumptions. 

Nevertheless, negative embedding space derived of these data will necessarily reflect more than just integrative opportunities or unobserved gaps between existing research topics. 
When examining individual documents, we noted that publications located in the negative conceptual space between topics were often interdisciplinary, integrating those topics.
However, negative space might also reflect a lack of expertise within a given field or subfield or at a given institution, given that universities tend to maintain (and intentionally curate) research specializations \cite{Braunerhjelm01062008, Pastor_Serrano_2016}. 
We recognize that performing this analysis on an exhaustive selection of publications from a single, smaller subfield would reduce opportunity for confounding by administrative factors. 
However, a corollary to this is that mixup might be utilized in aid of understanding what is `missing' from a research institute, university, department, or funding mechanism.
It is widely recognized that faculty hires are subject to biases, especially by race and sex \cite{Finkelstein2016, Weeden2017}.
This hiring process can be broken down into four phases, during which bias can affect decision-making: (1) forming a search committee and determining the specialty of the prospective hire; (2) recruitment and marketing the position; (3) evaluating the candidates; (4) final decisions \cite{OMeara_et_al_2020}. 
Assuming the goal is to identify candidates who offer a complementary expertise and skillset \cite{SMITH2021101104}, this metric might offer unique insights subject to fewer human biases when tailoring job searches, determining recruitment strategies, and evaluating candidates for "fit". 

This similarly applies to research funding agencies: identify negative space in a given field, or among their own funded projects, decide if that negative space is meaningful (a divide between astrophysics and social science is to be expected, and would not merit `filling'), then solicit appropriate applications.
Alternatively, within grants intended to promote interdisciplinary research and determine `unmet needs' (e.g., translational science institutes funded by the National Institutes of Health's National Center for Advancing Translational Sciences), this metric could be used to identify research areas that might merit further development. 
Ultimately, persistent homology and mixup represent an important opportunity to further understand conceptual landscapes and, by extension, better understand the embeddings underpinning a rapidly growing number of language models.

\subsection{Funding}
Research reported in this publication was supported by the University of Florida Clinical and Translational Science Institute, which is supported in part by the NIH National Center for Advancing Translational Sciences under award number UL1TR001427. 
The content is solely the responsibility of the authors and does not necessarily represent the official views of the National Institutes of Health.

\clearpage
\bibliographystyle{plain}
\bibliography{main.bib}

\begin{thebibliography}{10}

\bibitem{aenlle2009estrogen}
Kristina~K Aenlle, Ashok Kumar, Li~Cui, Travis~C Jackson, and Thomas~C Foster.
\newblock Estrogen effects on cognition and hippocampal transcription in middle-aged mice.
\newblock {\em Neurobiology of aging}, 30(6):932--945, 2009.

\bibitem{Andersonetal2004}
Neil Anderson, Carsten K.~W. De~Dreu, and Bernard~A. Nijstad.
\newblock The routinization of innovation research: a constructively critical review of the state-of-the-science.
\newblock {\em Journal of Organizational Behavior}, 25(2):147--173, 2004.

\bibitem{angelov2020}
Dimo Angelov.
\newblock Top2vec: Distributed representations of topics, 2020.

\bibitem{ARIA2017959}
Massimo Aria and Corrado Cuccurullo.
\newblock bibliometrix: An r-tool for comprehensive science mapping analysis.
\newblock {\em Journal of Informetrics}, 11(4):959--975, 2017.

\bibitem{Arseniev-Koehler2024}
Alina Arseniev-Koehler.
\newblock Theoretical foundations and limits of word embeddings: What types of meaning can they capture?
\newblock {\em Sociological Methods \& Research}, 53(4):1753--1793, 2024.

\bibitem{Bornmann_et_al_2016}
Lutz Bornmann, Andreas Thor, Werner Marx, and Hermann Schier.
\newblock The application of bibliometrics to research evaluation in the humanities and social sciences: An exploratory study using normalized google scholar data for the publications of a research institute.
\newblock {\em Journal of the Association for Information Science and Technology}, 67(11):2778--2789, 2016.

\bibitem{Borsuk1948}
Karol Borsuk.
\newblock On the imbedding of systems of compacta in simplicial complexes.
\newblock {\em Fundamenta Mathematicae}, 35(1):217--234, 1948.

\bibitem{botnan2020decomposition}
Magnus Botnan and William Crawley-Boevey.
\newblock Decomposition of persistence modules.
\newblock {\em Proceedings of the American Mathematical Society}, 148(11):4581--4596, 2020.

\bibitem{Braunerhjelm01062008}
Pontus Braunerhjelm.
\newblock Specialization of regions and universities: The new versus the old.
\newblock {\em Industry and Innovation}, 15(3):253--275, 2008.

\bibitem{Campelloetal2015}
Ricardo J. G.~B. Campello, Davoud Moulavi, Arthur Zimek, and J\"{o}rg Sander.
\newblock Hierarchical density estimates for data clustering, visualization, and outlier detection.
\newblock {\em ACM Trans. Knowl. Discov. Data}, 10(1), July 2015.

\bibitem{Chen2017140}
Chaomei Chen.
\newblock Science mapping: A systematic review of the literature.
\newblock {\em Journal of Data and Information Science}, 2(2):1--40, 2017.

\bibitem{Corcoran04032023}
Padraig Corcoran and Christopher~B. Jones.
\newblock Topological data analysis for geographical information science using persistent homology.
\newblock {\em International Journal of Geographical Information Science}, 37(3):712--745, 2023.

\bibitem{saussure1983course}
Ferdinand de~Saussure.
\newblock {\em Course in General Linguistics}.
\newblock Duckworth, London, 1983.
\newblock Translated by Roy Harris.

\bibitem{Draganov_2024}
Ond{\v r}ej Draganov and Steven Skiena.
\newblock The shape of word embeddings: Quantifying non-isometry with topological data analysis.
\newblock {\em Findings of the Association for Computational Linguistics: EMNLP 2024, pages 12080--12099. ACL}, 04 2024.

\bibitem{Diaz-Faes_et_al_2016}
Adrián~A. Díaz-Faes, María Bordons, and Thed~N. van Leeuwen.
\newblock Integrating metrics to measure research performance in social sciences and humanities: The case of the spanish csic.
\newblock {\em Research Evaluation}, 25(4):451--460, 10 2016.

\bibitem{edelsbrunner2010computational}
Herbert Edelsbrunner and John Harer.
\newblock {\em Computational topology: an introduction}.
\newblock American Mathematical Soc., 2010.

\bibitem{elsayed2022polygenic}
Abdelrahman~H Elsayed, Xueyuan Cao, Amit~K Mitra, Huiyun Wu, Susana Raimondi, Christopher Cogle, Zeina Al-Mansour, Raul~C Ribeiro, Alan Gamis, Edward~Anders Kolb, et~al.
\newblock Polygenic ara-c response score identifies pediatric patients with acute myeloid leukemia in need of chemotherapy augmentation.
\newblock {\em Journal of Clinical Oncology}, 40(7):772--783, 2022.

\bibitem{Feng_2021}
Michelle Feng and Mason~A. Porter.
\newblock Persistent homology of geospatial data: A case study with voting.
\newblock {\em SIAM Review}, 63(1):67--99, January 2021.

\bibitem{Finkelstein2016}
Martin~J. Finkelstein, Valerie~M. Conley, and Jack~H. Schuster.
\newblock Taking the measure of faculty diversity.
\newblock Technical report, TIAA Institute, 2016.
\newblock Accessed October 2025.

\bibitem{FlemingEtAl2007}
Lee Fleming, Santiago Mingo, and David Chen.
\newblock Collaborative brokerage, generative creativity, and creative success.
\newblock {\em Administrative Science Quarterly}, 52(3):443--475, 2007.

\bibitem{Fortunatoetal2018}
Santo Fortunato, Carl~T. Bergstrom, Katy Börner, James~A. Evans, Dirk Helbing, Staša Milojević, Alexander~M. Petersen, Filippo Radicchi, Roberta Sinatra, Brian Uzzi, Alessandro Vespignani, Ludo Waltman, Dashun Wang, and Albert-László Barabási.
\newblock Science of science.
\newblock {\em Science}, 359(6379):eaao0185, 2018.

\bibitem{gholizadeh2020}
Shafie Gholizadeh, Armin Seyeditabari, and Wlodek Zadrozny.
\newblock A novel method of extracting topological features from word embeddings, 2020.

\bibitem{DONUT}
Barbara Giunti, J{\=a}nis Lazovskis, and Bastian Rieck.
\newblock {DONUT}: {D}atabase of {O}riginal \& {N}on-{T}heoretical {U}ses of {T}opology, 2022.
\newblock \url{https://donut.topology.rocks}.

\bibitem{grootendorst2022}
Maarten Grootendorst.
\newblock Bertopic: Neural topic modeling with a class-based tf-idf procedure, 2022.

\bibitem{Hickok_2024}
Abigail Hickok, Benjamin Jarman, Michael Johnson, Jiajie Luo, and Mason~A. Porter.
\newblock Persistent homology for resource coverage: A case study of access to polling sites.
\newblock {\em SIAM Review}, 66(3):481--500, May 2024.

\bibitem{Hiraoka_2016}
Yasuaki Hiraoka, Takenobu Nakamura, Akihiko Hirata, Emerson~G. Escolar, Kaname Matsue, and Yasumasa Nishiura.
\newblock Hierarchical structures of amorphous solids characterized by persistent homology.
\newblock {\em Proceedings of the National Academy of Sciences}, 113(26):7035--7040, June 2016.

\bibitem{Hofstra_et_al_2020}
Bas Hofstra, Vivek~V. Kulkarni, Sebastian Munoz-Najar Galvez, Bryan He, Dan Jurafsky, and Daniel~A. McFarland.
\newblock The diversity–innovation paradox in science.
\newblock {\em Proceedings of the National Academy of Sciences}, 117(17):9284--9291, 2020.

\bibitem{Krenz2023tilltnet}
Till Krenz.
\newblock {\em tilltnet/rdimensions}.
\newblock None, 9 2023.

\bibitem{Landhuis_2016}
Esther Landhuis.
\newblock Scientific literature: Information overload.
\newblock {\em Nature}, 535(7612):457–458, Jul 2016.

\bibitem{le2014}
Quoc~V. Le and Tomas Mikolov.
\newblock Distributed representations of sentences and documents, 2014.

\bibitem{Leahey2023}
Erin Leahey, Jina Lee, and Russell~J. Funk.
\newblock What types of novelty are most disruptive?
\newblock {\em American Sociological Review}, 88(3):562--597, 2023.

\bibitem{Leahey&Moody2014}
Erin Leahey and James Moody.
\newblock Sociological innovation through subfield integration.
\newblock {\em Social Currents}, 1(3):228--256, 2014.

\bibitem{leroy2023resolution}
Victoria Leroy, Jun Cai, Zhenxiao Tu, Alexander McQuiston, Simrun Sharma, Amir Emtiazjoo, Carl Atkinson, Gilbert~R Upchurch~Jr, and Ashish~K Sharma.
\newblock Resolution of post-lung transplant ischemia-reperfusion injury is modulated via resolvin d1-fpr2 and maresin 1-lgr6 signaling.
\newblock {\em The Journal of Heart and Lung Transplantation}, 42(5):562--574, 2023.

\bibitem{li2021novel}
Changhai Li, Kunal Patel, Zhenxiao Tu, Xiaofeng Yang, Liudmila Kulik, Ali Alawieh, Patterson Allen, Qi~Cheng, Caroline Wallace, Jane Kilkenny, et~al.
\newblock A novel injury site-natural antibody targeted complement inhibitor protects against lung transplant injury.
\newblock {\em American Journal of Transplantation}, 21(6):2067--2078, 2021.

\bibitem{li2007chibby}
Feng-Qian Li, Amar~M Singh, Adaobi Mofunanya, Damon Love, Naohiro Terada, Randall~T Moon, and Ken-Ichi Takemaru.
\newblock Chibby promotes adipocyte differentiation through inhibition of $\beta$-catenin signaling.
\newblock {\em Molecular and Cellular Biology}, 27(12):4347--4354, 2007.

\bibitem{liu2007computational}
Tian Liu, Xueli Liu, Yunmei Chen, and Rongling Wu.
\newblock A computational model for functional mapping of genes that regulate intra-cellular circadian rhythms.
\newblock {\em Theoretical Biology and Medical Modelling}, 4(1):5, 2007.

\bibitem{marrero2023samhd1}
Richard~J Marrero, Xueyuan Cao, Huiyun Wu, Abdelrahman~H Elsayed, Jeffery~M Klco, Raul~C Ribeiro, Jeffrey~E Rubnitz, Xiaotu Ma, Soheil Meshinchi, Richard Aplenc, et~al.
\newblock Samhd1 single nucleotide polymorphisms impact outcome in children with newly diagnosed acute myeloid leukemia.
\newblock {\em Blood advances}, 7(11):2538--2550, 2023.

\bibitem{MARTIN20121219}
Ben~R. Martin.
\newblock The evolution of science policy and innovation studies.
\newblock {\em Research Policy}, 41(7):1219--1239, 2012.
\newblock Exploring the Emerging Knowledge Base of 'The Knowledge Society'.

\bibitem{McInnes2017}
Leland McInnes, John Healy, and Steve Astels.
\newblock hdbscan: Hierarchical density based clustering.
\newblock {\em Journal of Open Source Software}, 2(11):205, 2017.

\bibitem{McInnes2020}
Leland McInnes, John Healy, and James Melville.
\newblock Umap: Uniform manifold approximation and projection for dimension reduction, 2020.

\bibitem{munch2017user}
Elizabeth Munch.
\newblock A user’s guide to topological data analysis.
\newblock {\em Journal of Learning Analytics}, 4(2):47--61, 2017.

\bibitem{munkres2018elements}
James~R Munkres.
\newblock {\em Elements of algebraic topology}.
\newblock CRC press, 2018.

\bibitem{OMeara_et_al_2020}
KerryAnn O’Meara, Dawn Culpepper, and Lindsey~L. Templeton.
\newblock Nudging toward diversity: Applying behavioral design to faculty hiring.
\newblock {\em Review of Educational Research}, 90(3):311--348, 2020.

\bibitem{Parish_et_al_2018}
Austin~J. Parish, Kevin~W. Boyack, and John P.~A. Ioannidis.
\newblock Dynamics of co-authorship and productivity across different fields of scientific research.
\newblock {\em PLOS ONE}, 13(1):1--12, 01 2018.

\bibitem{Pastor_Serrano_2016}
José~Manuel Pastor and Lorenzo Serrano.
\newblock The determinants of the research output of universities: Specialization, quality and inefficiencies.
\newblock {\em Scientometrics}, 109(2):1255–1281, Aug 2016.

\bibitem{Perea2023}
Jose~A. Perea, Luis Scoccola, and Christopher~J. Tralie.
\newblock Dreimac: Dimensionality reduction with eilenberg-maclane coordinates.
\newblock {\em Journal of Open Source Software}, 8(91):5791, 2023.

\bibitem{pitman1937significance}
Edwin~JG Pitman.
\newblock Significance tests which may be applied to samples from any populations.
\newblock {\em Supplement to the Journal of the Royal Statistical Society}, 4(1):119--130, 1937.

\bibitem{Salton_68}
Gerard. Salton.
\newblock {\em Automatic Information Organization and Retrieval.}
\newblock McGraw Hill Text, 1968.

\bibitem{sharma2022lipoxin}
Ashish~Kumar Sharma, Jun Cai, Victoria Leroy, Zhenxiao Tu, Alejandro Gonzalez, Joseph Hartman, Jennifer Mulligan, Carl Atkinson, and Gilbert Upchurch~Jr.
\newblock Lipoxin a4 mitigates ferroptosis via fpr2 signaling during lung ischemia-reperfusion injury.
\newblock {\em The Journal of Immunology}, 208(Supplement\_1):175--13, 2022.

\bibitem{SinatraEtAl2016}
Roberta Sinatra, Dashun Wang, Pierre Deville, Chaoming Song, and Albert-László Barabási.
\newblock Quantifying the evolution of individual scientific impact.
\newblock {\em Science}, 354(6312):aaf5239, 2016.

\bibitem{SMITH2021101104}
Thomas~Bryan Smith, Raffaele Vacca, Till Krenz, and Christopher McCarty.
\newblock Great minds think alike, or do they often differ? research topic overlap and the formation of scientific teams.
\newblock {\em Journal of Informetrics}, 15(1):101104, 2021.

\bibitem{Smith_Vacca_Mantegazza_Capua_2021}
Thomas~Bryan Smith, Raffaele Vacca, Luca Mantegazza, and Ilaria Capua.
\newblock Natural language processing and network analysis provide novel insights on policy and scientific discourse around sustainable development goals.
\newblock {\em Scientific Reports}, 11(1), Nov 2021.

\bibitem{SutherlandEtAl2011}
William~J. Sutherland, David Goulson, Simon~G. Potts, and Lynn~V. Dicks.
\newblock Quantifying the impact and relevance of scientific research.
\newblock {\em PLOS ONE}, 6(11):1--10, 11 2011.

\bibitem{Thelwall_et_al_2015}
Mike Thelwall and Maria~M Delgado.
\newblock Arts and humanities research evaluation: no metrics please, just data.
\newblock {\em Journal of Documentation}, 71(4):817--833, 07 2015.

\bibitem{Thomas_2021}
Ashleigh Thomas, Kathleen Bates, Alex Elchesen, Iryna Hartsock, Hang Lu, and Peter Bubenik.
\newblock Topological data analysis of c. elegans locomotion and behavior.
\newblock {\em Frontiers in Artificial Intelligence}, 4, June 2021.

\bibitem{Tshitoyan_et_al_2019}
Vahe Tshitoyan, John Dagdelen, Leigh Weston, Alexander Dunn, Ziqin Rong, Olga Kononova, Kristin~A. Persson, Gerbrand Ceder, and Anubhav Jain.
\newblock Unsupervised word embeddings capture latent knowledge from materials science literature.
\newblock {\em Nature}, 571(7763):95–98, Jul 2019.

\bibitem{Uchendu_2024}
Adaku Uchendu and Thai Le.
\newblock Unveiling topological structures in text: A comprehensive survey of topological data analysis applications in nlp, 2024.

\bibitem{Ver_Hoef_2023}
Lander Ver~Hoef, Henry Adams, Emily~J. King, and Imme Ebert-Uphoff.
\newblock A primer on topological data analysis to support image analysis tasks in environmental science.
\newblock {\em Artificial Intelligence for the Earth Systems}, 2(1), January 2023.

\bibitem{Wagner:2024aa}
Hubert Wagner, Nickolas Arustamyan, Matthew Wheeler, and Peter Bubenik.
\newblock Mixup barcodes: Quantifying geometric-topological interactions between point clouds, 2024.

\bibitem{Wang_2011}
Bei Wang, Brian Summa, Valerio Pascucci, and Mikael Vejdemo-Johansson.
\newblock Branching and circular features in high dimensional data.
\newblock {\em IEEE Transactions on Visualization and Computer Graphics}, 17(12):1902--1911, 2011.

\bibitem{Weeden2017}
Kim~A. Weeden, Sarah Thébaud, and Dafna Gelbgiser.
\newblock Degrees of difference: Gender segregation of u.s. doctorates by field and program prestige.
\newblock {\em Sociological Science}, 4(6):123--150, 2017.

\bibitem{Weitzman1998}
Martin~L. Weitzman.
\newblock Recombinant growth*.
\newblock {\em The Quarterly Journal of Economics}, 113(2):331--360, 05 1998.

\bibitem{wu2004polyethylenimine}
Ke~Wu, Craig~A Meyers, Jennifer~A Bennett, Michael~A King, Edwin~M Meyer, and Jeffrey~A Hughes.
\newblock Polyethylenimine-mediated ngf gene delivery protects transected septal cholinergic neurons.
\newblock {\em Brain research}, 1008(2):284--287, 2004.

\bibitem{Wu_Wang_Evans_2019}
Lingfei Wu, Dashun Wang, and James~A. Evans.
\newblock Large teams develop and small teams disrupt science and technology.
\newblock {\em Nature}, 566(7744):378–382, Feb 2019.

\bibitem{xiong2005derivation}
Chen Xiong, Chang-Qing Xie, Li~Zhang, Jifeng Zhang, Kefeng Xu, Mingui Fu, Winston~E Thompson, Li-Jun Yang, and Yuqing~E Chen.
\newblock Derivation of adipocytes from human embryonic stem cells.
\newblock {\em Stem cells and development}, 14(6):671--675, 2005.

\bibitem{Cufar2020}
Matija Čufar.
\newblock Ripserer.jl: flexible and efficient persistent homology computation in julia.
\newblock {\em Journal of Open Source Software}, 5(54):2614, 2020.

\end{thebibliography}

\end{document}